\newcommand{\be}[0]{\begin{equation}}
\newcommand{\ee}[0]{\end{equation}}
\newcommand{\ba}[0]{\begin{eqnarray}}
\newcommand{\ea}[0]{\end{eqnarray}}
\newcommand{\nn}[0]{\nonumber}
\documentstyle[12pt,epsfig]{article}

\begin{document}
\begin{flushright}
\large
DTP-98/96\\
December 1998\\
\end{flushright}

\vspace{1.2cm}
\begin{center}
\Large
\centerline{\bf Determination of Radiative Widths of Scalar Mesons}
\vspace{0.3cm}

\centerline{\bf from Experimental Results on $\gamma\gamma\to\pi\pi$}
\vspace{1.5cm}
\large

M. Boglione $^{1}$

\vspace{0.5cm}
and
\vspace{0.5cm}

\large

M.R. Pennington $^2$

\vspace{0.4cm}
\begin{em}
$^1$  Vrije Universiteit Amsterdam, De Boelelaan 1081,\\ 1081 HV Amsterdam, 
The Netherlands\\

\vspace{0.4cm}
$^2$ Centre for Particle Theory, University of Durham\\
     Durham DH1 3LE, U.K.\\

\end{em}

\vspace{1.5cm}

\end{center}
\normalsize

{\leftskip = 2cm 
 \rightskip = 2cm
 \noindent
The scalar mesons in the 1 GeV region constitute the Higgs sector of the
strong interactions. They are responsible for the masses of all light
flavour hadrons. However, the composition of these scalar states is far
from clear, despite decades of experimental effort. The two photon couplings
of the $f_{0}$'s are a guide to their structure. Two photon results from Mark II, Crystal Ball and CELLO
prompt a new Amplitude Analysis  of $\gamma\gamma\to\pi^+\pi^-$,
$\pi^0\pi^0$ cross-sections. Despite their currently limited angular coverage and lack of polarized photons, we use a methodology that provides the nearest one can presently achieve to
a model-independent partial wave separation.
We find two distinct classes of solutions.  Both have very similar
two photon couplings for the $f_0(980)$ and $f_0(400-1200)$.
Hopefully these definitive results will be a spur to dynamical calculations
that will bring us a better understanding of these important states.
\par}
\newpage
\baselineskip=7.5mm
\parskip=2mm

\vspace{0.3cm}

\noindent
\section{Introduction}
\noindent 
Two photon processes are a remarkably useful tool for studying the structure
of matter and determining the composition of hadrons~\cite{reviews}.  Photons clearly couple to charged
 objects
and the observed cross-sections are directly related to these charges. Thus, for example,
 in the
reaction $\gamma\gamma\to\pi\pi$, the shape
of the integrated cross-sections 
perfectly illustrates this dynamics. At low energies, the photon
sees the pion as a whole entity and couples to
its charge. Consequently, the cross-section for $\gamma
\gamma \to \pi^{+} \pi^{-}$ is large at threshold,
whereas the $\gamma \gamma \to \pi^{0} \pi^{0}$ cross-section is very small~\cite{whalley}. 
When the energy increases, the shortening of its
wavelength enables the photon to see the individual constituents
of the pion, couples to their charges and causes them to resonate
(see, for instance,~\cite{mikedafne}). Both
the charged and neutral cross-sections are then dominated by the
Breit-Wigner peak corresponding to the $f_2(1270)$ resonance, with several underlying
$f_0$ states. The coupling of each of these to $\gamma\gamma$ is a measure
of the charges of their constituents (to the fourth power) and so helps to
build up a picture of the inner nature of  these mesons.
But how do we determine their $\gamma\gamma$ couplings from experimental data~?

\noindent In an ideal world, with complete information on all the possible angular 
correlations between the initial and final state directions and spins,
we could decompose
the cross-sections into components with definite sets of quantum numbers. 
From these, we could then unambiguously deduce the couplings to two photons of
all the resonances with those quantum numbers, not only the
$f_2(1270)$ but also the more complicated scalar resonances $f_0(980)$ 
and $f_0(400-1200)$ (and at higher energies the $f_0(1500)$ and $f_J(1710)$)~\cite{PDG}.
 Unfortunately, in the real world, experiments have only a
limited angular coverage and the polarization of the initial
state is not measured. This lack of information plays a crucial
role in any analysis and affects the determination of the resonance 
couplings~\cite{morpen88}. Thus one has to make assumptions
of varying degree of rigour : for instance, in the $f_2(1270)$ region,
assuming the cross-section is wholly $I=0$ $D$--wave with helicity two~\cite{hel2}. Estimates of the
underlying $I=0$ scalar couplings are made from the small $\pi^0\pi^0$ cross-section
at low energies, or the much larger $\pi^+\pi^-$ cross-section, etc~\cite{mars,boy}.  These are mere guesses and the
consequent results of doubtful certitude.

\noindent The aim of the present treatment is to perform an  Amplitude Analysis
in as model-independent way as possible.  To achieve this, we make up for our
lack of experimental information, firstly by analysing the charged ($\pi^+\pi^-$)
and neutral ($\pi^0\pi^0$) channels at the same time, and secondly using
severe theoretical constraints from Low's low energy theorem, crossing, analyticity and unitarity~\cite{morpen88,morpen90}.
The low energy theorem means that the amplitude for Compton scattering $\gamma\pi\to\gamma\pi$
is specified precisely at threshold~\cite{low}.  Analyticity, together with the fact that
the pion is so much lighter than any other hadron, means the Born amplitude,
modified in a  calculable way, dominates the $\gamma\gamma\to\pi\pi$
process in the near threshold regime~\cite{morpenpl,mikedafne}.  This provides the anchor on to which to hook
our Amplitude Analysis, determining all the partial waves with both $I=0, 2$ 
below 5 or 600 MeV. Above this energy, unitarity adds further constraints.
Each $\gamma\gamma\to\pi\pi$
partial wave amplitude is related to the corresponding hadronic processes
$h\to\pi\pi$.  Below 1 GeV or so, $h$ can only be $\pi\pi$ and the
constraints are highly restrictive.  Above 1 GeV, $K{\overline K}$ channels
not only open, but open strongly. This means we must incorporate 
coupled channel unitarity and include inputs from $\pi\pi\to K{\overline K}$ too.
Above 1.1 GeV, the $\eta\eta$ channel opens and above 1.4 GeV a series of multi-pion
channels become increasingly important. Because the $\eta\eta$ threshold
is relatively weak, and the $K{\overline K}$ channel the major source of 
inelasticity, we can reliably perform an Amplitude Analysis,
incorporating just  $\pi\pi$ and $K{\overline K}$ information up to 1.4 GeV or so.
Above that energy, we would have to access information on $\gamma\gamma\to n\pi$
and $\pi\pi\to n\pi$ (with $n \ge 4$) too and the analysis becomes impracticable,
at present.

\noindent Since 1990, when the last amplitude analysis of $\gamma \gamma \to \pi \pi$ was 
performed~\cite{morpen90}, new results on $\gamma \gamma \to 
\pi ^+ \pi ^-$ from the CELLO collaboration~\cite{beh}, more detailed
information on the scalar $\pi \pi$ final state interactions and 
increased statistics in the Crystal Ball experiment~\cite{bien} on 
$\gamma \gamma \to \pi^{0} \pi^{0}$ have become available. These provide the
impetus for a new analysis. In addition, there has been much speculation about the
nature of the scalar states in this region, their relation to the lightest 
$q{\overline q}$ multiplet, to multiquark states and to glueball candidates~\cite{nonet,wisgur,nils,glue}.
In each case, their two photon width is a key parameter in this debate.
Consequently, we need to put what we presently know about such widths
on as a firm a foundation as possible. 
Hopefully, this will be a spur to two photon studies at CLEO, at LEP and at future $B$--factories.
Further, improvements in data should allow the $\gamma\gamma$ widths of the
$f_0(1500)$ and $f_J(1710)$ to be fixed too. With this as the long term aim,
the present analysis will be able to limit 
the number of possible
solutions previously found and  obtain more stringent information 
particularly on the scalar sector below 1.4~GeV.

\newpage
\baselineskip=7.4mm

\section{Formalism and parametrization}

The unpolarized cross-section for dipion production by two
real photons is given by the contributions of two helicity amplitudes
$M_{++}$ and $M_{+-}$ (the subscripts label the helicities of the
incoming photons)~\cite{reviews,morpen88} : 
\begin{equation}
\frac{d \sigma}{d \Omega} = \frac{1}{128 \pi ^2 s} \sqrt{1- 4m^2 _{\pi} /s}
\; \left[ \; |M_{++}|^2 + \, |M_{+-}|^2 \, \right] .
\end{equation}
These two helicity amplitudes can be decomposed into partial waves as
\begin{equation}
M_{++}(s,\theta,\phi) = e^2 \sqrt{16 \pi} \sum _{J \geq 0} F_{J0}(s) \;
Y_{J0}(\theta, \phi) \; ,
\end{equation}
\begin{equation}
M_{+-}(s,\theta,\phi) = e^2 \sqrt{16 \pi} \sum _{J \geq 2} F_{J2}(s) \;
Y_{J2}(\theta, \phi) \; .
\end{equation}
The partial waves $F_{J \lambda} \, (\lambda=0,2)$ are the quantities we want 
to determine.

\noindent As explained in the Introduction, such an Amplitude
Analysis is not possible without some theoretical input. 
The first key constraint is unitarity.  
This relates the process
of two photons producing some specific hadronic final state to the hadronic 
production
of these same final particles.  Thus, for each amplitude with definite 
spin $J$,
helicity $\lambda$ and isospin $I$, unitarity requires (as illustrated in
Fig.~1 for $h = \pi\pi$, $K{\overline K}$)
\be
{\rm Im}\, {\cal F} _{J\lambda} ^{I}(\gamma \gamma \to \pi \pi) \, =\,  
\sum _h \rho _h\ {\cal F} _{J\lambda} ^{I}(\gamma \gamma \to h) ^{*}\, 
{\cal T} _{J} ^{I}(h \to \pi \pi)\;, 
\ee 
\noindent where the sum is over all hadronic intermediate states $h$ that
are kinematically allowed; $\rho_h$ being the density of states for
each such channel. We have dropped any dependence the hadronic partial wave amplitude ${\cal T}^I_{J}$ may have on helicity, as we shall, in practice, only be concerned with spinless final and intermediate states.  Eq.~(4) is, of course, linear in the two photon 
amplitudes, ${\cal F}$. However, each hadronic amplitude 
${\cal T}^I_{J} (h \to \pi\pi)$ satisfies 
the non-linear unitarity relation:
\be
{\rm Im}\, {\cal T} _{J} ^{I}(h \to \pi \pi) \, =\,  
\sum _{h'} \rho _h\ {\cal T} _{J} ^{I}(h \to h') ^{*}\, 
{\cal T} _{J} ^{I}(h' \to \pi \pi)\;.
\ee

\noindent This equation means that Eq.~(4) is satisfied by~\cite{amp87}
\be
{\cal F} _{J\lambda} ^{I}(\gamma \gamma \to \pi \pi) \, =\,  \sum _h\
\overline \alpha _{h} ^{I,J\lambda}\ {\cal T} _{J} ^{I}(h \to \pi \pi)\;,
\ee
\noindent where the $\overline \alpha _{h} ^{I,J\lambda}$ are functions of $s$, which
are real above $\pi\pi$ threshold. Thus, unitarity
relates the $\gamma\gamma\to\pi\pi$ partial wave amplitudes to a sum over hadronic
amplitudes with the same $\pi\pi$ final state weighted by the {\it coupling},
 $\alpha$,
of $\gamma\gamma\
$ to each contributing hadronic channel.  Clearly, this constraint
is only useful when we have information on all of the accessible hadronic
channels.  This restricts the present analysis to two photon energies
below 1.4 GeV, where $\pi\pi$ and $K{\overline K}$ channels are essentially all
that are relevant, Fig.~1. Then
\be
{\cal F} _{J\lambda} ^{I}(\gamma \gamma \to \pi \pi) \, =\,  
\overline \alpha _{\pi} ^{I,J\lambda}\ {\cal T} _{J} ^{I}
(\pi\pi \to \pi \pi) \, +\,  \overline \alpha _{K} ^{I,J\lambda}\ 
{\cal T} _{J} ^{I}(K \overline K \to \pi \pi)
\;.
\ee

\begin{figure}[ht]
\vspace{-3mm}
\begin{center}
\mbox{~\epsfig{file=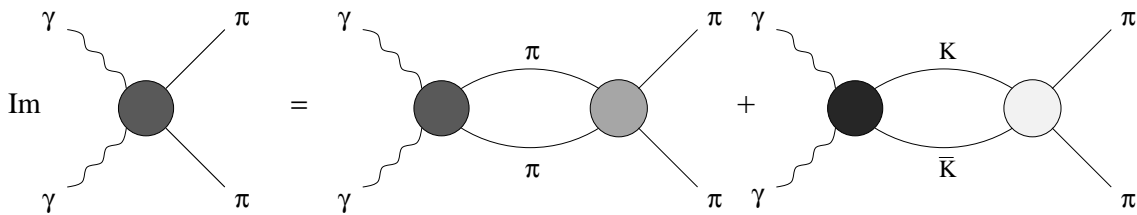,angle=-90,width=15cm}}
\vspace{3mm}
\caption{\leftskip 1cm\rightskip 6mm{The relation between the two photon amplitudes
and those for hadronic reactions given by Eq.~(7)}.}
\end{center}
\vspace{-4mm}
\end{figure}

\noindent$T$--invariance of the strong interactions means
$$
{\cal T} _{J} ^{I}(K \overline K \to \pi \pi)\, \equiv\,
{\cal T} _{J} ^{I}(\pi \pi \to K \overline K )\;.
$$
\noindent The analytic properties of the ${\cal F}^I_{J\lambda} (s)$ suggest the
functions ${\overline{\alpha}} (s)$ are smooth for $s \ge 4m_{\pi}^2$,
aside from possible poles that can occur in well-defined situations
that we will discuss in more detail below.  Notice that the
$\overline{\alpha}_{\pi,K}(s)$ give the weight with which
${\cal T}^I_J(\pi\pi\to\pi\pi)$ and ${\cal T}^I_J(\pi\pi\to K {\overline K})$,
respectively, contribute to ${\cal F}^I_{J\lambda}(\gamma\gamma\to\pi\pi)$.
The $\overline{\alpha}_{\pi}(s)$ and $\overline{\alpha}_{K}(s)$
will be determined by fitting the experimental data on $\gamma\gamma\to\pi\pi$,
as we describe in Sect.~4.

\noindent In Eq.~(7)  the hadronic amplitudes are independent of the
photon helicity $\lambda$, since the channels involve only
spinless pions and kaons.  Below 1 GeV, where the $K{\overline K}$ channel
switches off, Eq.~(6) can be expressed even more simply as
\be
{\cal F} _{J\lambda} ^{I}(\gamma \gamma \to \pi \pi)\, =\, 
\overline{a}  _{\pi} ^{I,J\lambda} \, {\cal T} _{J} ^{I}
(\pi\pi \to \pi \pi) 
\ee
\noindent where $\overline{a}$ is a real function for 
$4m_{\pi}^2 \le s \le 4m_{K}^2$.
Eq.~(7) and Eq.~(8) are, of course, consistent, since for
$4m_{\pi}^2 < s < 4m_K^2$ : 
\be
{\cal T}^I_J(\pi\pi\to K{\overline K})\, \propto\,
{\cal T}^I_J(\pi\pi\to\pi\pi)
\ee
 with a real function of proportionality.
The use of the representation Eq.~(7), throughout the region we consider, will
allow us to track through the important $K{\overline K}$ threshold region.
We stop at 1.4 GeV, since there multi-pion channels (as well as $\eta\eta$)
become increasingly important
and the unitarity constraint, Eq.~(6), more complicated and impossible to implement without
detailed partial wave information on $\pi\pi\to4\pi$, $6\pi$, etc.

\noindent Though unitarity imposes Eq.~(7), for each spin and isospin, 
in practice the $I=2$
amplitudes are simpler, as a result of the final state interactions being
weaker and the $K{\overline K}$ channel not being accessible. Consequently,
the representation, Eq.~(7), will only be used for the $I=0$ $S$ and $D$--waves. Let us deal with these in turn:
\begin{itemize}
\item
{\bf $I=0$ $S$--wave}: The $\gamma\gamma$ partial wave amplitude ${\cal F}^0_{00}$ will
be parametrized in terms of the two real coupling functions
$\overline{\alpha}_{\pi}^{0,00}$ and $\overline{\alpha}_{K}^{0,00}$
(denoted by $\overline{\alpha}_{\pi}^0$, $\overline{\alpha}_{K}^0$ as a shorthand) 
and the hadronic $S$--wave amplitudes ${\cal T}_S(\pi\pi\to\pi\pi)$ 
and ${\cal T}_S(\pi\pi\to K{\overline K})$.  The input for the hadronic amplitudes
is based on a modification (and extension) of the $K$--matrix parametrization of 
AMP~\cite{amp87}.
Briefly, the ${\bf T}$--matrix is related to the ${\bf K}$--matrix by
\be
{\bf {\cal T}} \, =\,  \frac{{\bf K}}{{\bf 1}\, -\, i{\bf \rho K}}
\label{K-def}
\ee
where ${\bf \rho}$ is the diagonal phase--space matrix. In the case in which  
two channels are considered, we have
\be
{\bf \rho} \, =\,  \left( 
\begin{array}{cc}
\rho _1 & 0 \\
0 & \rho _2
\end{array}
\right)\;,
\ee
where
\be
\rho _1 \, =\,  \sqrt{1\, -\, \frac{4m^2_{\pi}}{s}}\;,
\ee
\vspace{2mm}
\be
\rho _2 \, =\,  \frac{1}{2}\,\sqrt{1\, -\, \frac{4m^2_{K^{\pm}}}{s}} \, +\,  
          \frac{1}{2}\,\sqrt{1\, -\, \frac{4m^2_{K^{0}}}{s}}\;.
\ee
In this notation, the convention is
\[
1 \;\leftrightarrow \; \pi\pi \;,\quad
2 \;\leftrightarrow \; K \overline K.
\]
\noindent Coupled channel unitarity is then fulfilled by the $K$--matrix being real
for $s\ge 4m_{\pi}^2$.  It is then the $K$--matrix elements, $K_{ij}$
that embody the hadronic information. 
For our $I=0$ $S$--wave, we take the $K_1$ solution from ref.~\cite{amp87}
above approximately $1.1$ GeV, which is characterized by having only 
one pole in the ${\bf K}$--matrix in the $1$ GeV energy region.
For energies up to about $1$ GeV, we supply the strong interaction
amplitudes  ${\cal T}_S (\pi\pi \to \pi\pi)$ and 
${\cal T}_S (\pi\pi \to K \overline K)$ as given by the 
\lq\lq Solution~2\rq\rq ~obtained in a further refinement of the \cite{amp87} fit,  reported in 
ref.~\cite{revamp}. It includes a larger set of experimental data,
particularly in the $K\overline K$ threshold region, which favour a 
parametrization of the $f_0(980)$ that allows for two poles 
in the  
$T$--matrix on different sheets corresponding to this state. 
These solutions are smoothly joined from one energy region  to the other. 
The moduli of the amplitudes 
${\cal T} _S (\pi \pi \to \pi \pi)$ and ${\cal T} _S(\pi \pi \to K \overline K)$
determined in this way are shown in Fig.~1.
In fact, we have two $K$--matrix parametrizations of these, called ReVAMP1 and 2 ~\cite{revamp},
in which the $K$--matrix elements, in addition to having a single pole,
have either 2nd or 3rd order polynomials, respectively.
\begin{figure}[t]
\vspace{-0.6cm}
\begin{center}
\mbox{~\epsfig{file=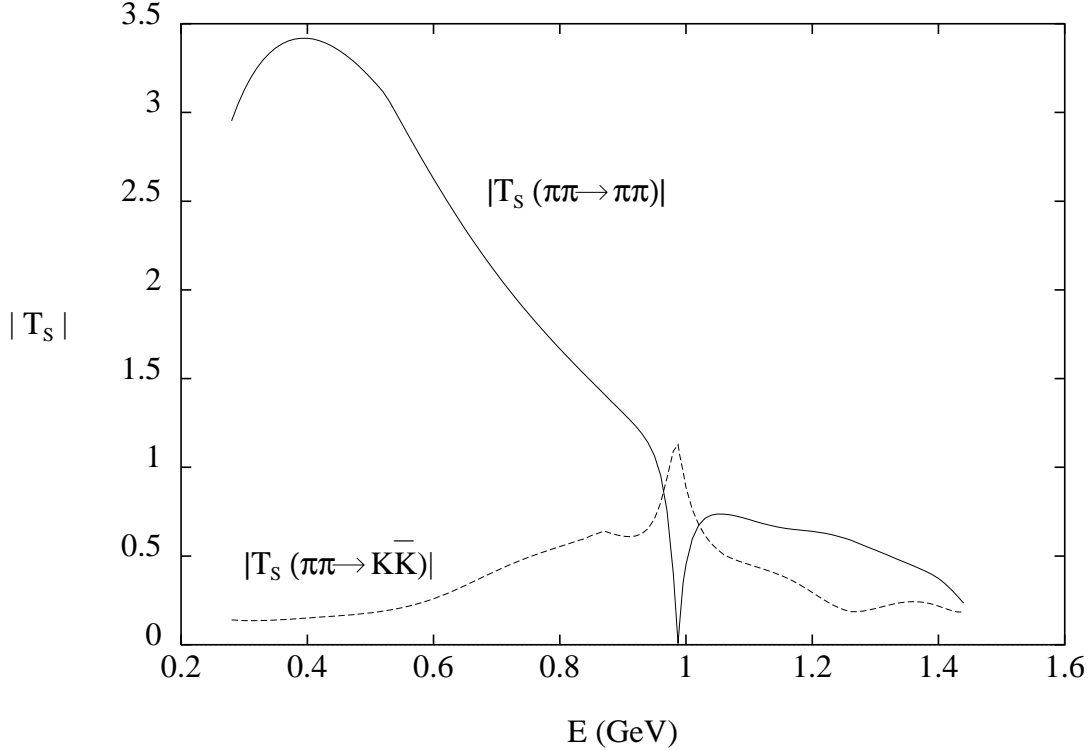,angle=-90,width=15.5cm}}
\caption{\leftskip 1.5cm\rightskip 0.8cm{Strong interaction inputs for the $I=0$ $S$--wave: here we show the
moduli of the amplitudes ${\rm T}$, which are obtained from the amplitudes 
${\cal T}$ by removing the Adler zero factor, 
${\rm T}(s)={\cal T}(s)/(s-s_0)$.}}
\end{center}
\vspace{-5mm}
\end{figure}

\noindent The relation of these hadronic amplitudes to that for $\gamma\gamma\to\pi\pi$
is through the coupling functions $\overline{\alpha}_{\pi,K}$.
Since these functions have only left hand cuts, they must be real for $s\ge 4m_{\pi}^2$,
but should be smooth along the right hand cut, once obvious dynamical structures are taken
into account. Such obvious structures are that the coupling functions have poles
wherever any element, or other sub-determinant, of the $T$--matrix has a {\bf real} zero.
For instance, ${\cal T}_S(\pi\pi\to\pi\pi)$ and ${\cal T}_S(\pi\pi\to K{\overline K})$ have Adler zeros below threshold
(at $s=s_0$), which are not present in the $\gamma\gamma$ amplitude.
(That is why they have been divided out in the amplitudes plotted in Fig.~2)
\footnote{ in general the $\pi\pi\to\pi\pi$ and $\pi\pi\to K{\overline K}$
amplitudes do not have their Adler zeros at exactly the same point $s=s_0$.
However, present data are not sensitive to small differences between their
positions and so in AMP \cite{amp87}, these were taken to be at the same point, 
for simplicity.}
Similarly, ${\rm{det}} {\bf{\cal T}}(s)$ can (and in fact does) vanish
at a real value of $s=s_1$, where  $s_1 \le 4m_K^2$ because of Eq.~(9).
Unless the $\overline{\alpha}$'s have poles at this point, with related residues, this 
constraint on the hadronic amplitudes unnecessarily transmits to 
the $\gamma\gamma$ amplitudes.
To avoid these unnatural constraints, the functions are parametrized as follows:
\ba
\nn
\overline {\alpha} ^0 _{\pi} (s) &=&  \frac{\alpha ^S _{\pi}(s)}{s-s_0} \,
 +\,  \frac{\nu}{s-s_1}\, {\cal T}_S (\pi\pi\to K{\overline K};s=s_1) \;,\nn \\
&& \nonumber\\
\overline {\alpha} ^0 _{K} (s) &=&  \frac{\alpha ^S _{K}(s)}{s-s_0} \, -\,  
\frac{\nu}{s-s_1}\, {\cal T}_S (\pi\pi\to\pi\pi;s=s_1)\;.
\ea
\noindent The fit then determines $\alpha_{\pi}^S$ and $\alpha_{K}^S$
as polynomials in $s$ and the constant parameter~$\nu$. 
It is important to stress that the poles in Eq.~(14)
do not appear in the $\gamma\gamma$ amplitudes, of course. This representation,
Eq.~(7),
for the $I=0$ $S$--wave has the flexibility needed to determine the details
of the mechanism by which the scalar resonances, $f_0$'s, couple to two photons.
\item
{\bf $I=0$ $D$--waves with $\lambda=0,2$}: here  a simplification arises from
the fact that the $\pi\pi\to\pi\pi$ and $\pi\pi\to K{\overline K}$ amplitudes
are proportional to each other, both being dominated by the $f_2(1270)$ resonance.
Then the hadronic amplitude ${\cal T}_D(\pi\pi\to\pi\pi)$ is given by
\be
{\cal T} _D (\pi\pi\to\pi\pi)\, =\, \frac{BR}{\beta(m_f^2)}\,
\frac{m_f\ \Gamma(s)}{m_f^2 - s - im_f \Gamma(s)} \;
\ee
where $BR$ is the branching ratio of $f_2\to\pi\pi$. Importantly, the width
is energy dependent and given by
\be
\Gamma(s)\, =\, \frac{\beta(s)}{\beta(m_f^2)}\, \Gamma_{\rm tot} \, D_2(s)
\ee
with $\beta(s) = \sqrt{1 - 4m_{\pi}^2/s}$. The factor
$D_2(s)$ incorporates threshold and barrier effects. Here, we take these
to be given by {\it duality shaping}, with the scale set by the slope of the 
non-strange Regge trajectories 
(or equivalently the $\rho$--mass, $m_{\rho}$). Then
\be
\nonumber
D_2(s)\,=\,{\left(1+\frac{6m_{\rho}^2}{(m_f^2-4m_{\pi}^2)}+
\frac{6m_{\rho}^4}{(m_f^2-4m_{\pi}^2)^2}\right)}/
{\left(1+\frac{6m_{\rho}^2}{(s-4m_{\pi}^2)}+
\frac{6m_{\rho}^4}{(s-4m_{\pi}^2)^2}\right)} \; .
\ee
The $\pi\pi\to K{\overline K}$ amplitude
is proportional to this and so (just as in Eq.~(8)) we can write
\be
{\cal F} _{2\lambda}\, =\,  \overline \alpha ^{0,2\lambda}(s)\, {\cal T} _D
(\pi\pi\to\pi\pi)\;, 
\label{D_0}
\ee
\noindent even above $K{\overline K}$ threshold, where the 
$\overline{\alpha}^{0,2\lambda}(s)$ are again smooth real functions of energy to be determined by the fit to experimental data.
To ensure the appropriate threshold behaviour for the $\gamma\gamma\to\pi\pi$
amplitude, these $D$--wave coupling functions are parametrized by modifying 
the threshold factor, so
\be
\overline \alpha ^{0,2 \lambda}(s) \, =\,  \frac{\alpha ^{D\lambda}}
{\sqrt{D_2(s)}}\;.
\ee
\noindent $m_f$ and $\Gamma_{\rm tot}$ are taken from the 
PDG Tables~\cite{PDG} and $D_2(s)$ as in Eq.~(17).
\end{itemize}

\baselineskip=7.4mm

\noindent So far the formalism we have described would apply to any reaction,
by which a non-strongly interacting initial state leads to $\pi\pi$.  We now turn to the
particular features of the two photon reaction.

\noindent For $\gamma \gamma \to \pi\pi$, Low's low energy theorem \cite{low} imposes an 
important constraint, in which the hadron charge fixes the size of the cross-section. 
This is embodied in the one pion exchange Born amplitude. Though the theorem 
applies at the threshold for the Compton process $\gamma \pi \to \gamma \pi$, 
the Born term controls the $\gamma \gamma \to \pi\pi$ amplitude in the whole 
low energy region~\cite{morpenpl}, as discussed extensively in \cite{mikedafne}.
It is this dominance of the Born term that means, unusually for a strong 
interaction
 final state, that the $I=2$ channel is just as important as that with $I=0$. 
 It is the almost exact cancellation between those amplitudes with $I=0,2$ that makes the  
$\gamma \gamma \to \pi ^0 \pi ^0$ cross-section small at low energies.

\noindent While the Born amplitude controls the low energy process, it is of course
 modified by final state interactions. As already mentioned these affect the $I=0$ 
 and $I=2$ $\pi\pi$ channels quite differently. For the $I=2$ final state,
  these interactions are weak and the Born amplitude is little changed, 
  remaining predominantly real  in all partial waves. In contrast, the $I=0$ 
  final state interactions in $S$ and $D$ waves are strong (leading to resonance 
  formation for instance), consequently even close to $\pi \pi$ threshold the Born 
  amplitude is modified. It is unitarity that allows these modifications to be 
  reliably calculated up to 600 or 700 MeV~\cite{morpenpl,mikedafne}. Consequently, the Born amplitude,
   with such modifications from final state interactions, provides a precise 
   description of the partial wave amplitudes on to which we must connect our 
   amplitude analysis.

\begin{itemize}
\item
{\bf All waves with spin $J \ge 4$}: For these,  final state interactions are negligible and so the $\gamma \gamma \to \pi\pi$ amplitudes are set equal to the Born amplitude in the whole energy region up to 1.4 GeV. Thus
\be
{\cal F}_{J \lambda}\,(\gamma \gamma \to \pi ^+ \pi ^-) \, =\,  
B_{J \lambda}\,(\gamma \gamma \to \pi ^+ \pi ^-) \;,
\ee
and
\be
{\cal F}_{J \lambda}\,(\gamma \gamma \to \pi ^0 \pi ^0) \, =\,  
B_{J \lambda}\,(\gamma \gamma \to \pi ^0 \pi ^0)\, = \, 0\;.
\ee
\vspace{0.01mm}
\item
{\bf $I=2$ $S$ and $D$ waves}: Here the $\gamma \gamma \to \pi \pi$ amplitudes have modifications from final state interactions  that can be calculated up to 1.4 GeV. This we do by expressing the amplitude essentially as a modulus times a phase factor as 
 \be
{\cal F}^{I=2}_{J\lambda} (s) \, =\,  \sqrt{\frac{1}{3}} \, P ^{B} _{J \lambda}
 (s)\, \Omega ^{I=2}_{J \lambda} (s)\;, 
\label{POmega}
\ee
where $\Omega ^{I=2}_{J \lambda} (s)$ is the appropriate Omn\`es function 
\cite{omnes}:
\be
\Omega _{J\lambda} ^{I} (s)\, =\,  \exp \left[ \frac{s}{\pi} \int _{4m^2_{\pi}} 
^{\infty} ds'\, \frac{ \phi  _{J\lambda} ^{I=2} (s')}{s'(s'-s)} \right]\;,
\ee
with $\phi  _{J\lambda} ^{I=2}$ the phase of the corresponding $\gamma \gamma$ 
amplitudes  ${\cal F}^{I=2}_{J \lambda}$. Applying elastic unitarity 
relates this phase to the $I=2$ spin $J$ $\pi\pi$ phase shift. 
Then, using the data of Ref. \cite{hoog}, the $\Omega$'s are readily computed. 
The function $P(s)$ in Eq. (\ref{POmega}) is then real for 
$s \ge 4 m_{\pi} ^2$ and can be calculated as follows.
Dropping the $I,\lambda$ indices to keep the notation simple, consider the analytic function $f_J (s)$ defined by
\be
f_J(s)\, = \, B_J(s) \, (\Omega _J ^{-1}(s) \, -\,  1)\;,
\ee
where $B_J(s)$ is the spin $J$ Born amplitude. We now write a once subtracted dispersion relation for $f_J(s)/(s-4m_{\pi} ^2)^{J/2}$
\ba
f_J(s)\, =\, f_J(0) &+& \frac{s(s-4m_{\pi} ^2)^{J/2}}{\pi}\, 
\int ^{\infty} _{4m_{\pi} ^2}ds' 
\frac{B_J(s')\,{\rm Im}\Omega _J ^{-1}(s')}{s'(s'-s)(s'-4m_{\pi} ^2)^{J/2}} 
\nonumber \\
&&\nonumber\\ 
&+& \frac{s(s-4m_{\pi} ^2)^{J/2}}{\pi}\, \int _{-\infty} ^{0}  
\frac{{\rm Im}[B_J (s') (\Omega _J ^{-1}(s')-1)]}{s'(s'-s)
(s'-4m_{\pi} ^2)^{J/2}}\;,
\label{disp.rel-f}
\ea
where we have taken into account that $B_J(s)$ has only a cut on the
left-hand side, so that
\be
\begin{array}{lll}
{\rm Im}\ f_J(s) \, =\, 
B_J(s)\,{\rm Im}\Omega _J ^{-1}(s) & \;\;{\rm for} & s \geq 4m_{\pi} ^2 \;, \\
\nonumber & & \\
{\rm Im}\ f_J(s) \, =\,  {\rm Im}\,[B_J(s) \,(\Omega _J  ^{-1}(s)-1)] & \;\;{\rm for} 
& s < 0 \;.
\end{array} 
\ee
If we use the fact that $f_J(0)=0$ since $\Omega (0)=1$, and subtract the
function $P(s)$ from Eq.~(\ref{disp.rel-f}), we
find
\be
P_J(s)\, =\, B_J(s)\, +\, \frac{s(s-4m_{\pi} ^2)^{J/2}}{\pi}\,
\int ^{\infty} _{4m_{\pi} ^2}ds' 
\frac{B_J(s')\,{\rm Im}\Omega _J ^{-1}(s')}{s'(s'-s)(s'-4m_{\pi} ^2)^{J/2}} 
\label{P-right-cut}
\ee 
which involves the integration of the imaginary part of the function
$\Omega _J ^{-1}(s)$ over the right--hand cut only. This allows us
to use hadronic information for $s \geq 4m_{\pi} ^2$ to constrain the input into 
our description of the limited
$\gamma\gamma\to\pi\pi$ experimental results. 

\noindent
The factor  $\sqrt{1/3}$ in Eq.~(\ref{POmega})
is the appropriate Clebsch--Gordan
coefficient, as obtained by decomposing the amplitudes with definite
isospin in terms of the amplitudes with definite charge quantum numbers.   
As a consequence of the fact that the $\pi ^0 \pi ^0$ Born amplitude 
is zero, we have:
\be
B (\gamma \gamma \to \pi ^0 \pi ^0 )\, =\, 
\sqrt{\frac{2}{3}} \, B ^{I=2} - \sqrt{\frac{1}{3}} \, B ^{I=0} = 0 \;,
\ee
from which
\be
B ^{I=0} \, =\,  \sqrt{2}\,B ^{I=2}\;.
\ee
Applying this to the Born amplitude
\be
{\cal B} \, \equiv \, B(\gamma \gamma \to \pi ^+ \pi ^-) \, =\, 
\sqrt{\frac{2}{3}} \, B ^{I=0} + \sqrt{\frac{1}{3}} \, B ^{I=2} 
\ee
we have
\be
B ^{I=2} \, =\,  \sqrt{\frac{1}{3}} \, {\cal B}\;\; , \;\;\; 
B ^{I=0} \, =\,  \sqrt{\frac{2}{3}} \, {\cal B}\;.
\ee
\vspace{2mm}
\item
{\bf $I=0$ $S$ and $D$ waves}: For these, an identical procedure can be used to 
calculate the modifications from final state interactions reliably up to 0.6 GeV, 
starting from 
\be
{\cal F}^{I=0}_{J\lambda} (s) \, =\,  \sqrt{\frac{2}{3}} \, P ^{B} _{J \lambda}
 (s)\, \Omega ^{I=0}_{J \lambda} (s)\;, 
\ee 
and using information about the $I=0$ $S$ and $D$ waves $\pi\pi$
 phase-shift~\cite{cm,mikedafne} to compute the $\Omega$'s. Below 600 MeV,
  the effect of the unknown $\gamma \gamma \to \pi \pi$ isoscalar phases
  in the inelastic regime above 1 GeV is small --- see Ref.~\cite{penn}.
The effect of the modification from final state interactions is shown for 
example in Fig.~3. At higher energies, non-pion exchange contributions become increasingly important as discussed in~\cite{mikedafne} and so 
these partial waves can no longer be reliably calculated from first principles. Instead, we leave the data to determine the amplitudes using the representation given by Eq.~(7).

\end{itemize}

\noindent
Let us summarize the input and the constraints, in the region under study.
 Everywhere, all the $I=2$ partial waves and the $I=0$ waves with $J \ge 4$ are
  given by the modified Born amplitudes. Final state interactions only appreciably affect the $I=0$ $S$ and $D$ waves.
The $I=0$ waves with $J=0,2$ can be reliably predicted by the modified Born amplitudes below 600 MeV. 
However, everywhere they can be represented by Eq. (7) which follows merely from coupled channel unitarity.

\noindent
With the hadronic amplitudes $\cal{T}_S$ and $\cal{T}_D$ known, 
the four coupling functions $\alpha ^S_{\pi}$, $\alpha ^S _{K}$,
 $\alpha ^{D\lambda}$ are what the data and the above constraints determine. 
 These four $\alpha$'s are polynomials in energy, which we allow to be at most cubic: this gives a reasonable
degree of flexibility, but without an overdue number of unphysical
structures that could affect the reliability of the fit. They are written 
as a Legendre expansion in terms of the variable $x$,
defined as
\be
x\, =\, \frac{2E - E_1 - E_2}{E_2 - E_1}\quad ,
\ee
so that the energy interval \{$E_1 = 0.28,  E_2 = 1.44 $\},
which gives the boundaries of the energy range we are fitting, maps onto the interval
$-1 \le x \le 1$. Symbolically, we write each $\alpha$ as 
\be
\alpha _{i} \, =\,  \sum _{n=0} ^{3} \alpha ^{(n)} _L P_n (x)\;.
\ee

\noindent 
Before we consider the data we are going to analyse to determine 
their partial wave content, let us stress  that there is an 
important region below 600 MeV, where the $\gamma \gamma \to \pi \pi$ 
amplitudes are predicted and must also agree with the unitary representation 
of Eq. (7). These predictions provide a reliable starting point
 for such a general representation. We next describe how  we build in this constraint.
\newpage
\baselineskip=7.2mm

\section{Analysis procedure}

\subsection{Low energy inputs}
To implement the constraint from Low's theorem~\cite{low}, that as we go down in energy 
the amplitudes are given by their Born terms~\cite{morpenpl}, we adopt the following strategy.

\begin{figure}[h]
\vspace{0.6cm}
\begin{center}
\mbox{~\epsfig{file=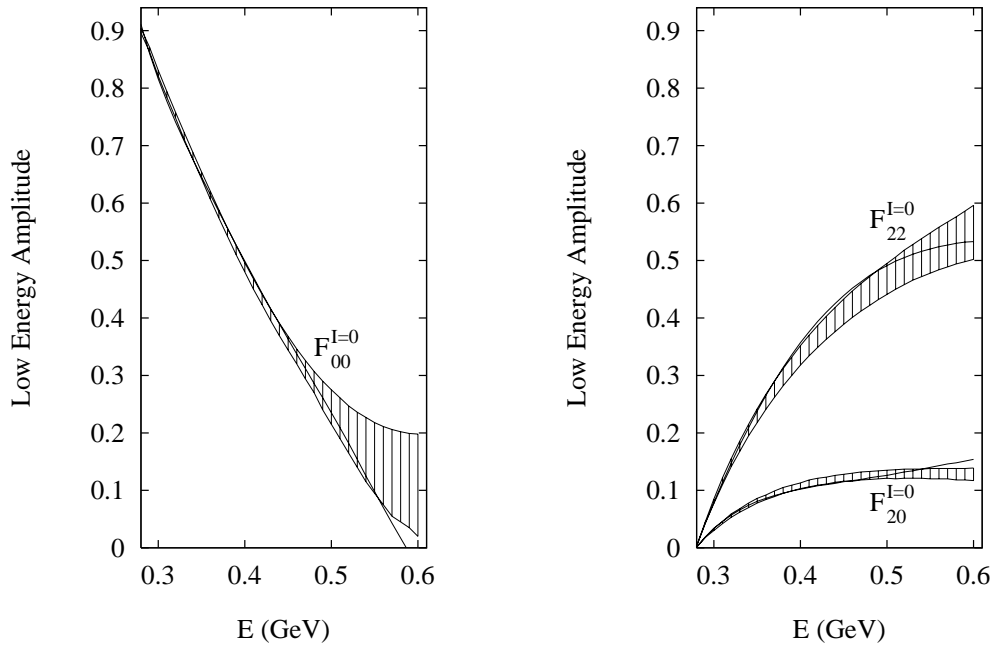,angle=-90,width=13cm}}
\vspace{-8.6cm}
\caption{\leftskip 0.4cm\rightskip 0.9cm{The amplitudes ${\cal F}^{I=0} _{J\lambda}$ in the low energy region.
The shaded areas represent the bands we give as a constraining input, by
building a \lq\lq horn\rq\rq around the central curve generated by the final
state modified Born amplitudes. The solid lines are the output of our fit.
For this plot we have used, as an illustration, the outcomes corresponding
to what is later called the {\it peak} solution; but the difference between the two solutions in the
low energy range is hardly noticable.}}
\end{center}
\end{figure}
\vspace{-.5cm}

\noindent We take the $[{\cal F}^{I=0} _{J\lambda}]_{\rm low \; energy}$ below
600 MeV, as calculated in Sect.~2. Our amplitudes, fitted to experiment, must agree with these within some tolerance. To fix this,  we construct a \lq\lq horn\rq\rq
~around the curves given by ${\cal F}^{I=0} _{J\lambda}(s)$, by assigning error bars,
 which are zero at threshold and become progressively larger as the energy increases. This reflects the fact that neglecting other exchanges then the pion becomes a poorer approximation as the energy increases. To do this we introduce a contribution to the total $\chi ^2$ from fitting
\be
\chi ^2 _{\rm low \; energy} \, =\,  \left( \frac{[{\cal F}^{I=0} _{J\lambda}]
_{\rm trial \; form} - [{\cal F}^{I=0} _{J\lambda}]_{\rm low \; energy}}
{\Delta {\cal F}} \right) ^2\;,
\label{low-en-x}
\ee
where $\Delta {\cal F}$ is obviously given by the errors on 
$[{\cal F}^{I=0} _{J\lambda}]_{\rm low \; energy}$ we have
introduced.
These ensure an extremely tight constraint close to $\pi\pi$ threshold,
while allowing larger flexibility when the two photon 
energy approaches $500-600$ MeV.
Fig.~3 illustrates how the amplitudes ${\cal F}^{I=0} _{J\lambda}$ 
determined by our fit fulfil the low energy constraints given by the Born
amplitudes modified by final state interactions of Eq.~(32).   
The shaded regions are the \lq\lq horns\rq\rq ~inside which the low energy
amplitudes ${\cal F}^{I=0} _{J\lambda}$ are required to fall.

\subsection{Data analysis}

\noindent
As mentioned in
the introduction, the data-sets on two photon scattering into charged pion
final states at low energies come from Mark II \cite{boy} and CELLO
\cite{beh}, whereas two different runs of the Crystal Ball experiment, the
last with much higher statistics,
provide the only available normalized experimental information on 
$\gamma\gamma \to \pi ^0\pi ^0$ for such low energies.
Table~1 shows the number of data in each experiment, below 1.4 GeV.

\begin{table}[t]
\begin{center}
\begin{tabular}{||c|c||c|c||c|c||}
\hline \hline
\rule[-0.4cm]{0cm}{12mm} Experiment & Process & 
 Int. X-sect. & $|\cos \theta |_{max}$ & Ang. distrib. 
 & $|\cos \theta |_{max}$ \\ 
\hline \hline
\rule[-0.8cm]{0cm}{20mm} 
Mark II & $\gamma \gamma \to \pi ^{+} \pi^{-}$ & 87 & 0.6 & 69 & 0.6 \\  
\hline 
\rule[-0.8cm]{0cm}{20mm}   
Cr. Ball & $\gamma \gamma \to \pi ^{0} \pi^{0}$ & 26 & 
\parbox{2.0cm}{0.8~(CB88) \protect \\ 0.7~(CB92)} & 80 & 0.8 \\
\hline
\rule[-0.8cm]{0cm}{20mm}   
CELLO & $\gamma \gamma \to \pi ^{+} \pi^{-}$ & 30 & 0.6 &  
\parbox{2.4cm}{127~(Harjes) \protect \\ 249~(Behrend)} & 0.55 - 0.8 \\  
\hline \hline 
\end{tabular}
\vspace{0.4cm}
\caption{\leftskip 0.4cm\rightskip 0.4cm{Number of data in each experiment
 below 1.4 GeV. Mark II results are from Boyer {\it et al.} \cite{boy},
 Crystal Ball from Marsiske {\it et al.} (CB88) \cite{mars} and Bienlein 
 {\it et al.} (CB92) \cite{bien}, 
 and CELLO from Harjes \cite{harjes} 
 and Behrend {\it et al.} \cite{beh}. }}
\end{center}
\end{table}
%
\noindent Though the angular distributions contain information about the integrated 
cross-section, because of different bin centres and sizes, these are not always the same. For instance, Mark II gives the angular distributions for 
$|\cos \theta| \le 0.6$  in energy bins of 100 MeV, but 
present the integrated cross-section in 10 MeV steps above 750 MeV.

\noindent From the CELLO experiment, we have angular distributions in 
$\Delta \cos \theta$ bins of 0.05 from Behrend {\it et al.} \cite{beh} and 
 $\Delta \cos \theta$ bins of 0.1 from the thesis of Harjes \cite{harjes}, both in energy bins of 50 MeV width.
Though these come from the same data sample we believe, we have fitted 
them as separate data-sets but weighted appropriately (see later),
 since the different binning produces quite a difference in the scatter 
 of the data-points, see Fig.~4.

\begin{figure}[h]
\vspace{9mm}
\begin{center}
\mbox{~\epsfig{file=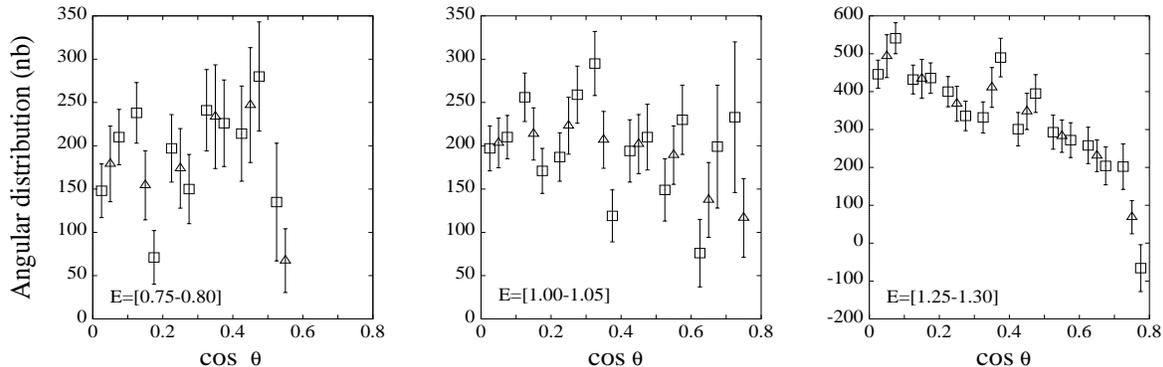,angle=-90,width=11cm}}
\vspace{-6.7cm}
\caption{\leftskip 0.7cm\rightskip 0.3cm{Comparison between the angular 
distribution of 
 CELLO data-points on $\gamma\gamma\to\pi^+\pi^-$ from Behrend \cite{beh} (squares) and Harjes \cite{harjes} (diamonds) for some 
 illustrative energy bins ($E$ is in GeV).}}
\end{center}
\vspace{-4mm}
\end{figure}

\noindent
Where statistical and systematic errors are quoted (see tables in 
\cite{whalley}),
these have been added in quadrature.
Each experiment has an absolute normalization for the cross-sections. 
However, these inevitably provide additional systematic uncertainties. Such uncertainties have been
 included in the results produced by the special low energy triggering of Mark II. However, above 700 MeV a systematic shift in normalization is apparent between the Mark II and CELLO integrated cross sections, though both of them are for 
$|\cos \theta |\le 0.6$, see Fig.~5.
It is clear that we must allow for some systematic shift in normalization,
 if we are to describe both data-sets in a sensible way.
 Mark II quote a systematic normalization uncertainty of 7\%.
 With this in mind, we allow for up  to a $5\%$ relative shift in normalization
  between Mark II and CELLO experiments.

\begin{figure}[p]
\begin{center}
\mbox{~\epsfig{file=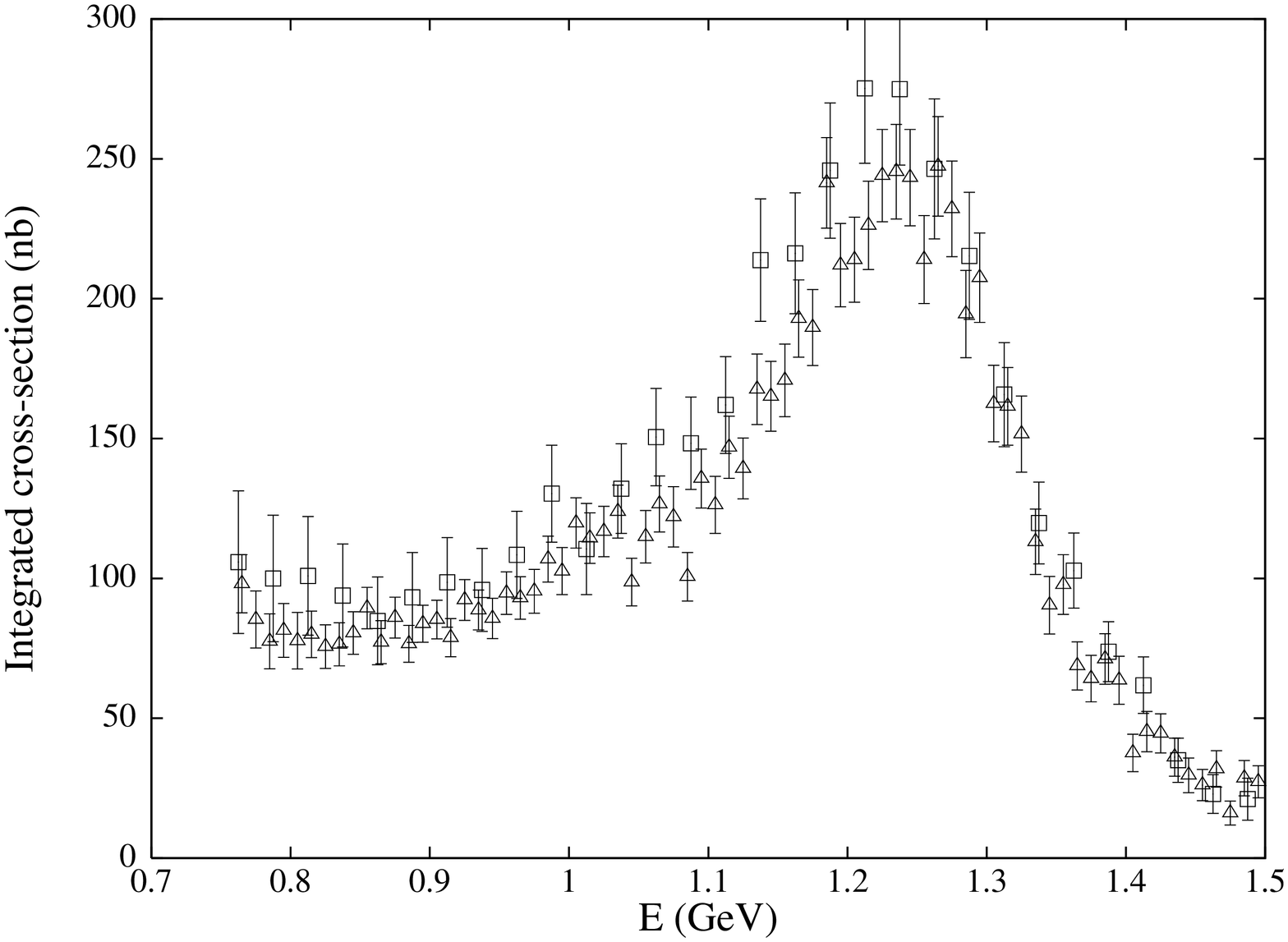,angle=-90,width=13.5cm}}
\vspace{0.5cm}
\caption{\leftskip 0.8cm\rightskip 0.6cm{Comparison between Mark II \cite{boy} (diamonds) and
 CELLO \cite{beh} (squares) integrated cross-sections for
$\gamma\gamma\to\pi^+\pi^-$ in their common energy range.}}
\end{center}

\begin{center}
\mbox{~\epsfig{file=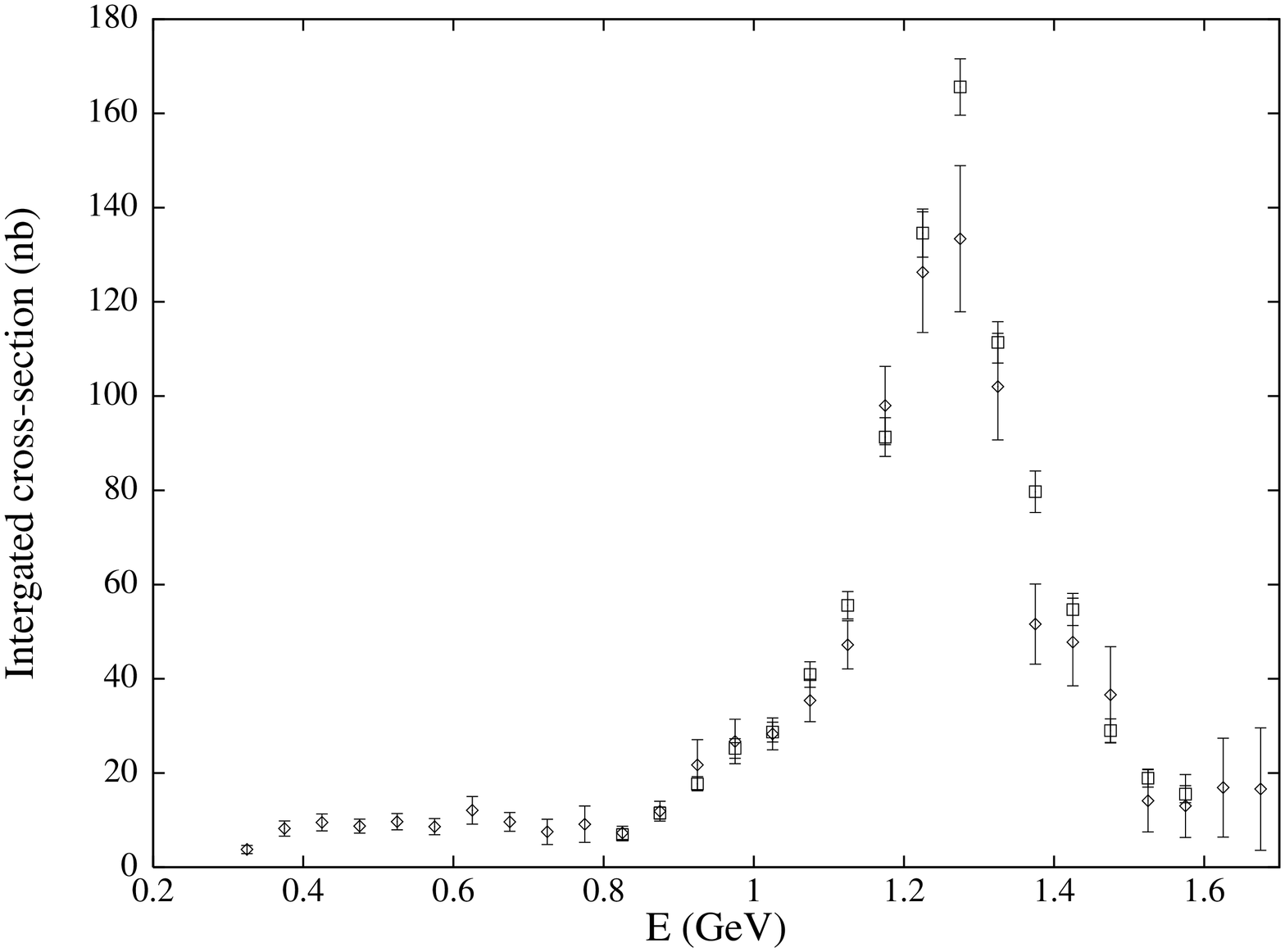,angle=-90,width=13.5cm}}
\vspace{0.5cm}
\caption{\leftskip 0.8cm\rightskip 0.6cm{Comparison between CB88 \cite{mars} 
(diamonds) and 
CB92 \cite{bien} (squares) integrated cross-sections
for $\gamma\gamma\to\pi^0\pi^0$.}}
\end{center}
\end{figure}
 
\noindent
For the $\pi ^0 \pi ^0$ channel, Crystal Ball had two distinct runs. The first covering the energy region from $\pi\pi$ threshold, called CB88~\cite{mars}. The second had 1.5 times as much data, but only above 800 MeV.
Bienlein {\it et al.}~\cite{bien}  combined this with the CB88 set to produce their complete CB92 dataset above 800 MeV. Nevertheless, there was always a clear systematic difference in the earlier and later runs through the $f_2$--region.
 The  CB88 set had a higher, narrower peak than the combined CB92 set 
 (see Fig.~6). 
 Since the CB88 set (above 800 MeV) is subsumed in CB92, 
 we cannot really separate them. 
 Consequently, we allow for a 
 systematic shift of $3\%$ between CB88 below 800 MeV and CB92 above.
 
 \noindent While both charged and neutral pion experiments would be consistent
 (and the fits even better) with larger normalization shifts, we have erred on the side of caution in allowing this freedom.
 To repeat : we assume that the results from CELLO, CB88 below 800 MeV and
  Mark II  below 400 MeV are absolute, and allow small shifts of other data-sets with respect to these.

\subsection{Weighting, relative normalizations and fitting}

\noindent
From Table 1, we see that the number of data-points is far greater 
for the reaction $\gamma \gamma \to \pi ^+ \pi ^-$ than for
 $\gamma \gamma \to \pi ^0 \pi ^0$. Since an accurate separation of 
 the $I=0$ component requires both are accurately described, we give
  different weight factors to each dataset. We choose these so that the
   Mark II and the CELLO data have roughly the same number of weighted data, 
   while the weight assigned to the Crystal Ball data approximately equals 
   the weighted sum of the Mark II and CELLO data. 
Nevertheless, good agreement is not easy to achieve.

\noindent The analysis program (GAMP~\cite{morpen90}) works by integrating the amplitudes
over the appropriate bin in energy and angle for each data-point. 
It  does not just use the
central values. This is to allow for any strong local variation of the amplitudes,
particularly near $K{\overline K}$ threshold. In the next Section, where we display the
solutions we find, this should be borne in mind.  Where the energy bins are sizable
(as with Crystal Ball), histograms are plotted (see Figs.~8 and 11). Where the energy bins are fine (as with Mark II data in
10 MeV steps), the fits are shown more appropriately as continuous lines joining the
bin centres, as in Figs.~7, 9, 10, 12. However throughout, the fits are histograms.
\newpage
\baselineskip=7.2mm

\section{Results}

\noindent
Our fits deliver two classes of distinct solutions.
As is seen from Figs.~7-18, the two classes have very similar quality
as far as fitting the data is concerned, yet they have quite distinct
characteristics. The first 
has a peak in the cross--section
located in the $1$ GeV energy region and from now on will be referred to as
the {\it peak} solution. The second has a dip
 in the same energy region and will be called the {\it dip} solution.
The plots with the fits to the Mark II $\pi^+\pi^-$ integrated cross-section
look more structured than those of CELLO in the 1 GeV region (cf. Figs~7 and 9, or 10 and 12). This is because our fitting routine integrates over bins, which for Mark II are only 10 MeV wide in this region compared with 25 MeV from CELLO. Detailed dynamical features are picked up more strongly in the finer binned data. Recall the fits are  not continuous curves but histograms, as shown in Figs.~8, 11 for the $\pi^0\pi^0$ data of Crystal Ball.

\noindent
 Looking at the plots of just the Mark II results, Figs~7,~10,
 on the integrated cross-section for $\gamma\gamma\to\pi^+\pi^-$
 in the 1 GeV region, one might be tempted to conclude that
 the {\it peak} solution is disfavoured. However, one cannot conclude that
 from the CELLO results, Figs.~9,~12, on the same channel and Crystal Ball,
Figs.~8,~11, on the $\pi^0\pi^0$ final state, 
 nor indeed from looking at the fits over the whole energy region for Mark II.
 Thus, individual features are a poor guide, even though one's eye naturally picks those out.
   Indeed, the overall quality of the fits in each sector for the
 two distinct solutions are quite comparable as can be seen from Tables~2
 and 3. There 
we report the contributions to the total $\chi^2$ from
each experimental set, for the integrated cross-section and the angular
distribution separately for the best of these solutions.
This total is expressed as the $\chi^2$ per degree of freedom. 
The {\it peak} solution has a slightly lower overall 
$\chi ^2$. However, the CELLO data turn out to be easiest to fit (and so we 
have reduced their weight by 1/2 to achieve greater sensitivity to the rest). 
In contrast, the two solutions have appreciably greater $\chi ^2$ for the 
Mark II and Crystal Ball results by $\sim 0.5$ per degree of freedom.

\begin{table}[p]
\begin{center}
\begin{tabular}{||c|c||c|c|c|c||}
\hline \hline
\multicolumn{6}{||c||}{\rule[-0.4cm]{0cm}{12mm} PEAK SOLUTION
\parbox{1.2cm}{~~~~} $\chi ^2 _{{\rm tot}} = 1.40$} \\
\hline
\rule[-0.4cm]{0cm}{12mm} Experiment & Process & data-points & 
$\chi ^2_{{\rm average}}$ & Int. X-sect.  & Ang. distrib. \\ 
\hline \hline
\rule[-0.8cm]{0cm}{20mm} 
Mark II & $\gamma \gamma \to \pi ^{+} \pi^{-}$ & 156 & 1.54 &
$\chi ^2$ = 1.82 & $\chi ^2$ = 1.19 \\  
\hline 
\rule[-0.8cm]{0cm}{20mm}   
Cr. Ball & $\gamma \gamma \to \pi ^{0} \pi^{0}$ & 106 & 1.44  & 
$\chi ^2$ = 1.42 & $\chi ^2$ = 1.44 \\  
\hline
\rule[-1.0cm]{0cm}{22mm}   
CELLO & $\gamma \gamma \to \pi ^{+} \pi^{-}$ & 406 & 1.33 & 
$\chi ^2$ = 0.65 & 
\parbox{2.6cm}{$\chi ^2$~=1.13 \protect \\  from~Harjes \protect \\  
$\chi ^2$~=1.52 \protect \\ from Behrend} \\
\hline \hline 
\end{tabular}
\vspace{0.4cm}
\caption{\leftskip 0.5cm\rightskip 0.5cm{Summary of contributions from each experiment to the total $\chi
^2$ for the {\it peak} solution. Here $\chi^2 _{\rm tot}$ is calculated by
dividing the sum of the  $\chi^2$'s for all data-sets by the total number
of data-points we are fitting, namely 668. $\chi^2 _{\rm average}$ is
computed by dividing the sum of the  $\chi^2$'s for each data-set
by the total number of data-points in that
experiment. }}
\end{center}
\begin{center}
\begin{tabular}{||c|c||c|c|c|c||}
\hline \hline
\multicolumn{6}{||c||}{\rule[-0.4cm]{0cm}{12mm} DIP SOLUTION
\parbox{1.2cm}{~~~~} $\chi ^2 _{{\rm tot}} = 1.48$} \\
\hline
\rule[-0.4cm]{0cm}{12mm} Experiment & Process & data-points & 
$\chi ^2_{{\rm average}}$ & Int. X-sect.  & Ang. distrib. \\ 
\hline \hline
\rule[-0.8cm]{0cm}{20mm} 
Mark II & $\gamma \gamma \to \pi ^{+} \pi^{-}$ & 156 & 1.75 &
$\chi ^2$ = 1.99 & $\chi ^2$ = 1.46 \\  
\hline 
\rule[-0.8cm]{0cm}{20mm}   
Cr. Ball & $\gamma \gamma \to \pi ^{0} \pi^{0}$ & 106 & 1.62  & 
$\chi ^2$ = 1.97 & $\chi ^2$ = 1.51 \\  
\hline
\rule[-1.0cm]{0cm}{22mm}   
CELLO & $\gamma \gamma \to \pi ^{+} \pi^{-}$ & 406 & 1.33 & 
$\chi ^2$ = 0.88 & 
\parbox{2.6cm}{$\chi ^2$~=1.09 \protect \\  from~Harjes \protect \\  
$\chi ^2$~=1.51 \protect \\ from Behrend} \\
\hline \hline 
\end{tabular}
\vspace{0.4cm}
\caption{\leftskip 1cm\rightskip 1cm{Summary of contributions from each experiment to the total $\chi
^2$ for the {\it dip} solution, as described for Table~2.}}
\end{center}
\end{table}

\begin{figure}[p]
\begin{center}
\vspace{-1.2cm}
~\epsfig{file=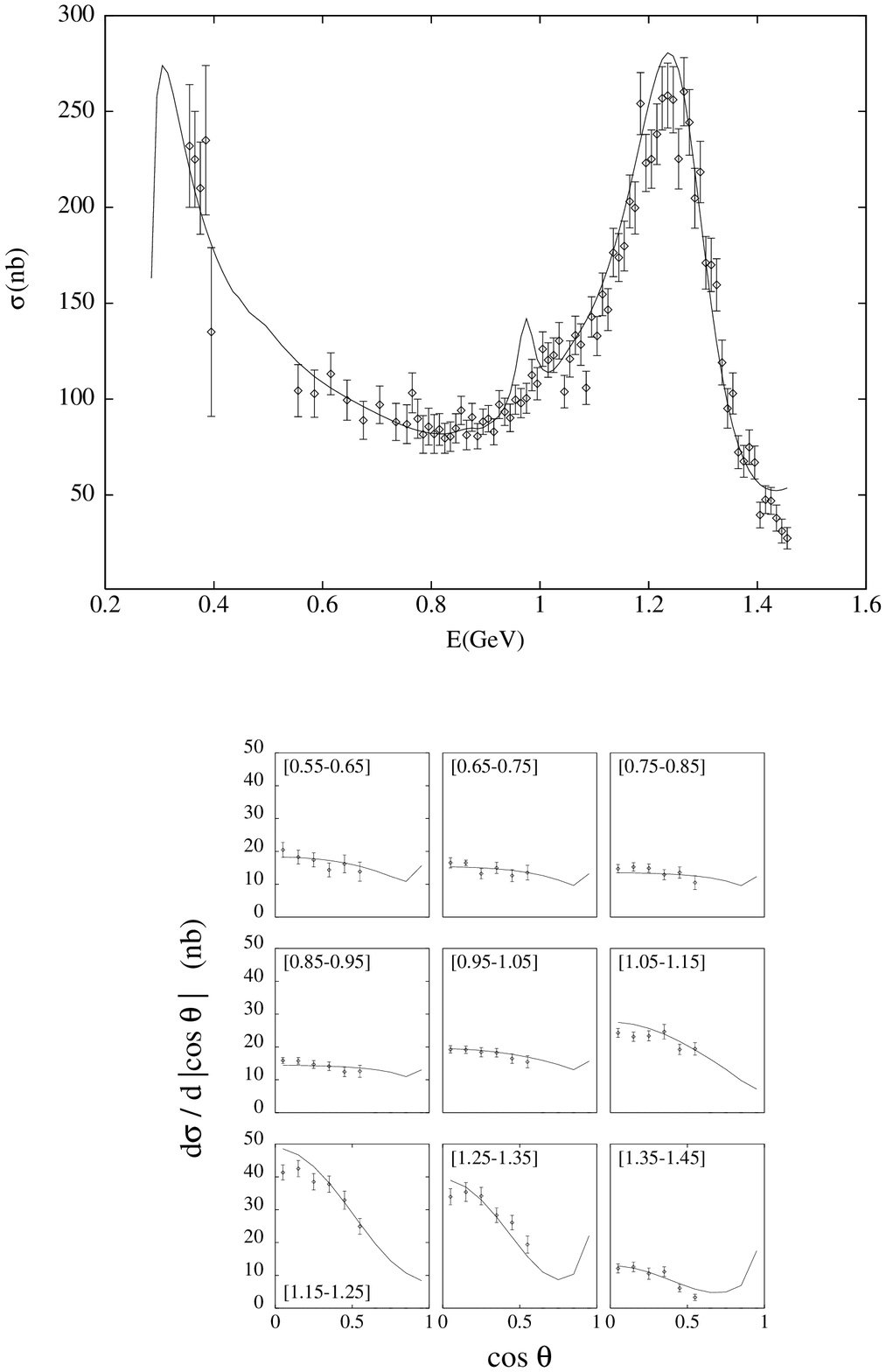,angle=0,width=16.0cm}
\vspace{-2.5cm}
\caption{\leftskip 1.5cm\rightskip 0.5cm{Cross-section as a function of the
 $\pi \pi$ invariant mass 
integrated over $\mid \cos\ \theta \mid \; \leq \; 0.6$, and 
angular distributions as a function of $\cos\ \theta$, 
for the $\gamma \gamma \rightarrow \pi^+\pi^-$ process, from the MARK II
experiment \cite{boy} ({\it peak} solution).}}
\end{center}
\end{figure}

\begin{figure}[p]
\begin{center}
\vspace{-0.5cm}
~\epsfig{file=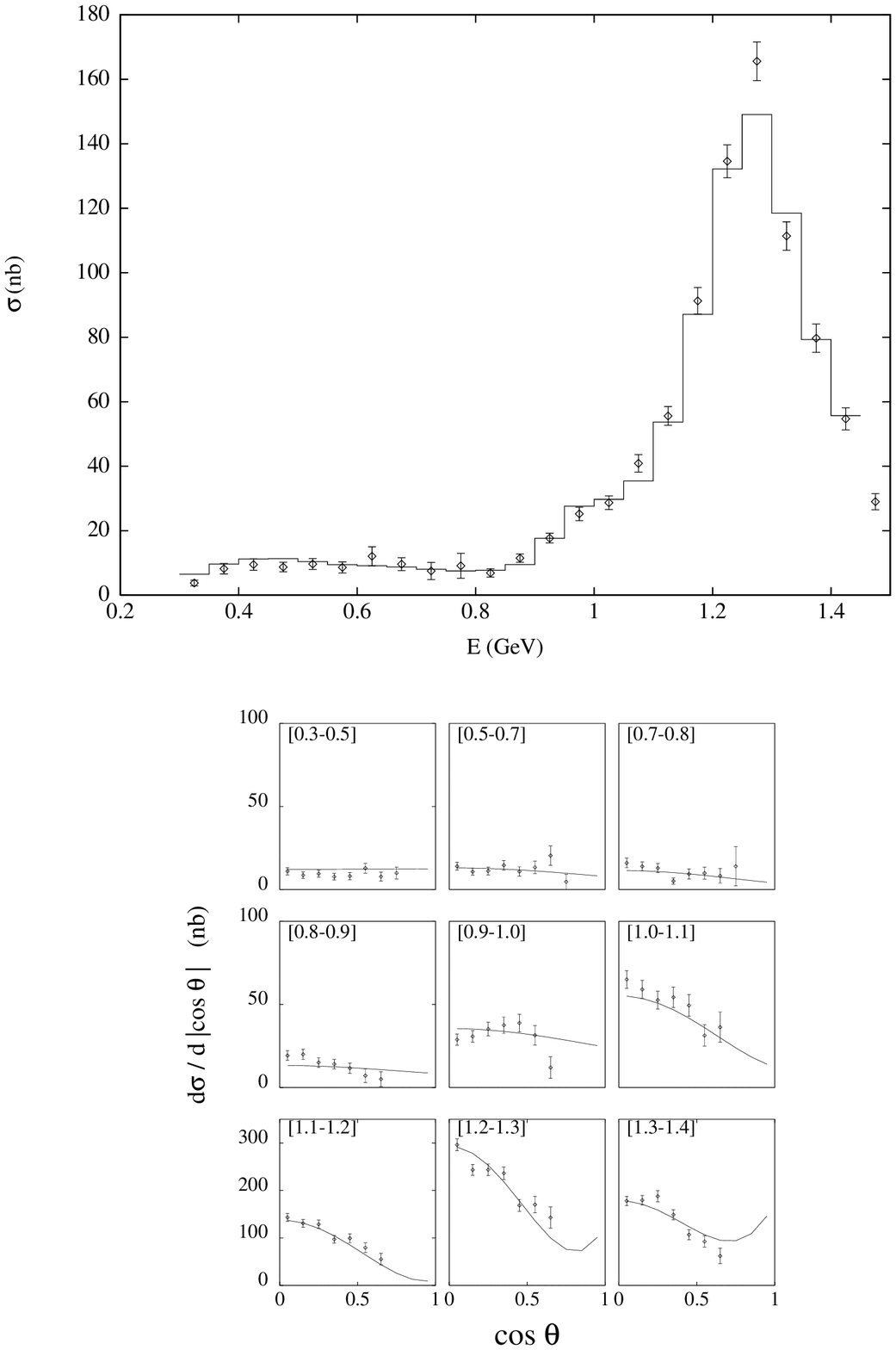,angle=0,width=16.5cm}
\vspace{-2.5cm}
\caption{\leftskip 1.5cm\rightskip 0.5cm{Cross-section as a function of the $\pi \pi$ invariant mass 
integrated over $\mid \cos\ \theta \mid \; \leq \; 0.8$ for $E \, \leq \,
0.8 \,$GeV and $\mid \cos\ \theta \mid \; \leq \; 0.7$ for higher energies, and 
angular distributions as a function of $\cos\ \theta$, 
for the $\gamma \gamma \rightarrow \pi^0 \pi^0$ process, from the CRYSTAL BALL
experiment \cite{mars,bien} ({\it peak} solution).}}
\end{center}
\end{figure}

\begin{figure}[p]
\begin{center}
\vspace{-1.5cm}
~\epsfig{file=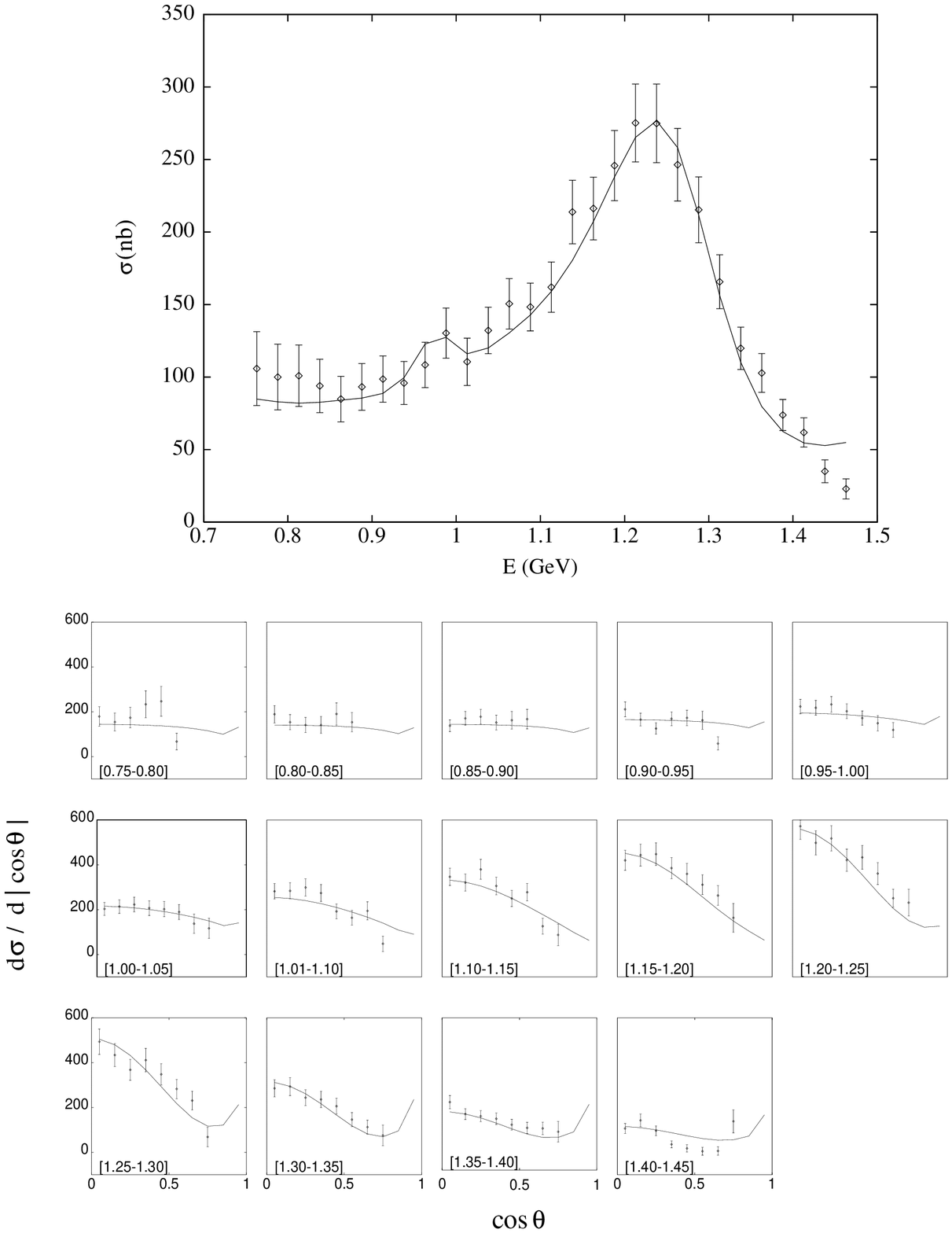,angle=0,width=13.5cm}
\caption{\leftskip 0.1cm\rightskip 1cm{Cross-section as a function of the $\pi \pi$ invariant mass 
integrated over $\mid \cos\ \theta \mid \; \leq \; 0.6$, and 
angular distributions as a function of $\cos\ \theta$, 
for the $\gamma \gamma \rightarrow \pi^+ \pi^-$ process, from the CELLO
experiment ({\it peak} solution). The angular distributions are from the
binning of Harjes~\cite{harjes} --- see Fig.~13 for the Behrend {\it et al.} binning.}}
\end{center}
\end{figure}

\begin{figure}[p]
\begin{center}
\vspace{-1.2cm}
~\epsfig{file=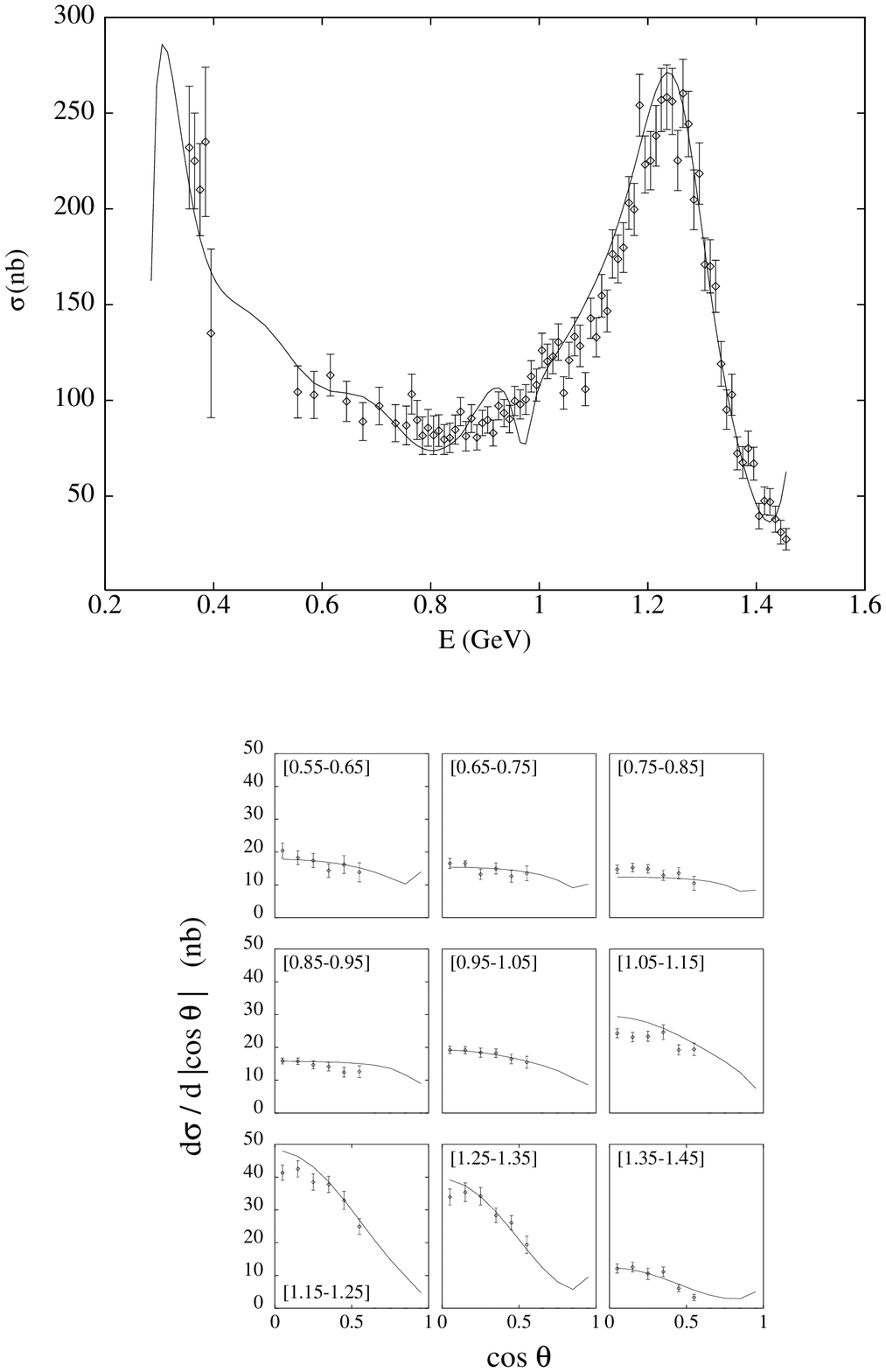,angle=0,width=16.cm}
\vspace{-2.5cm}
\caption{\leftskip 1.5cm\rightskip 0.5cm{Cross-section as a function of the
 $\pi \pi$ invariant mass 
integrated over $\mid \cos\ \theta \mid \; \leq \; 0.6$, and 
angular distributions as a function of $\cos\ \theta$, 
for the $\gamma \gamma \rightarrow \pi^+\pi^-$ process, from the MARK II
experiment \cite{boy} ({\it dip} solution).}}
\end{center}
\end{figure}

\begin{figure}[p]
\begin{center}
\vspace{-0.5cm}
~\epsfig{file=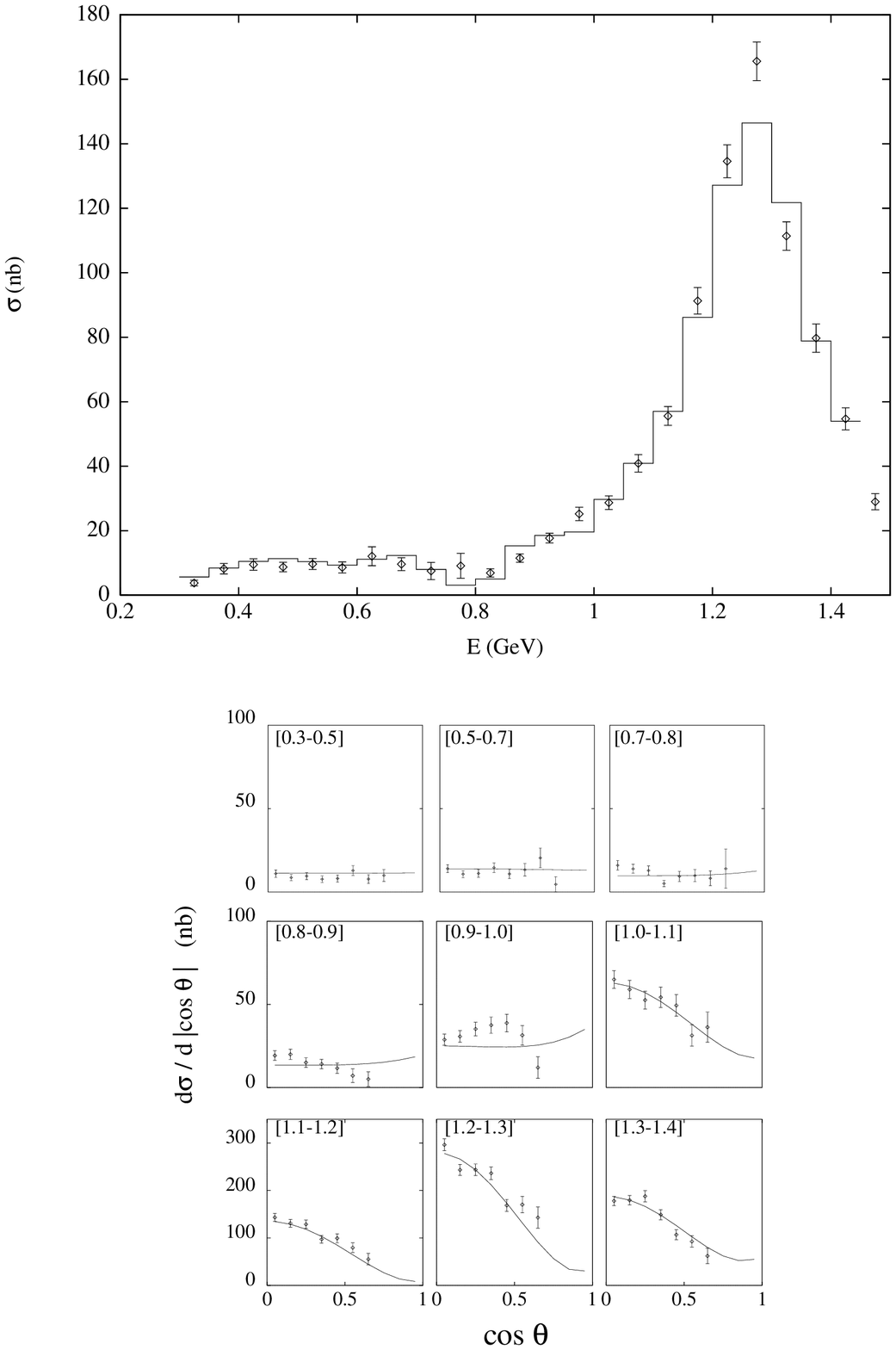,angle=0,width=16.5cm}
\vspace{-2.5cm}
\caption{\leftskip 1.5cm\rightskip 0.5cm{Cross-section as a function of the $\pi \pi$ invariant mass 
integrated over $\mid \cos\ \theta \mid \; \leq \; 0.8$ for $E \, \leq \,
0.8 \,$GeV and $\mid \cos\ \theta \mid \; \leq \; 0.7$ for higher energies, and 
angular distributions as a function of $\cos\ \theta$, 
for the $\gamma \gamma \rightarrow \pi^0 \pi^0$ process, from the CRYSTAL BALL
experiment \cite{mars,bien} ({\it dip} solution).}}
\end{center}
\end{figure}

\begin{figure}[p]
\begin{center}
\vspace{-1.5cm}
~\epsfig{file=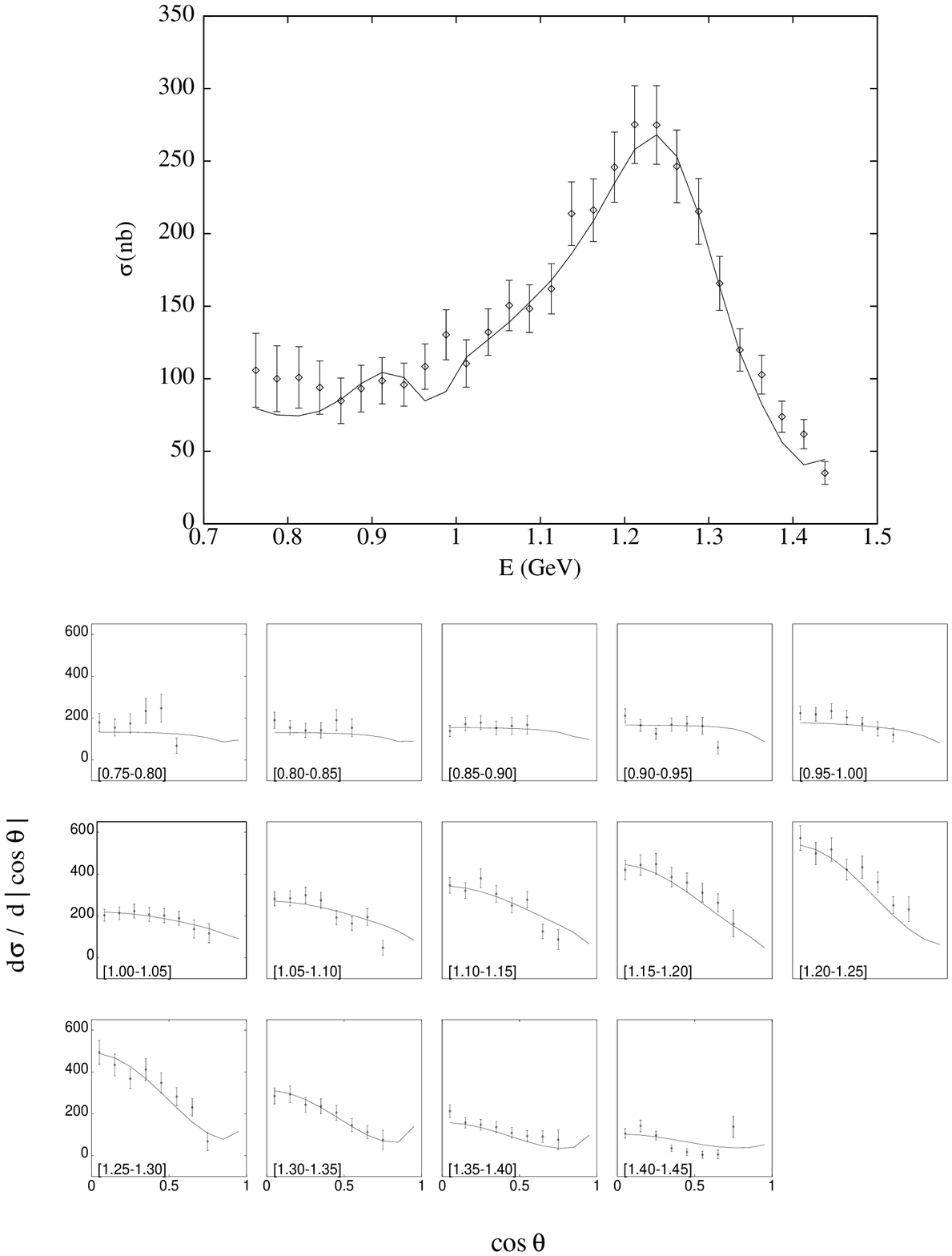,angle=0,width=13.5cm}
\caption{\leftskip 0.1cm\rightskip 1cm{Cross-section as a function of the $\pi \pi$ invariant mass 
integrated over $\mid \cos\ \theta \mid \; \leq \; 0.6$, and 
angular distributions as a function of $\cos\ \theta$, 
for the $\gamma \gamma \rightarrow \pi^+ \pi^-$ process, from the CELLO
experiment ({\it dip} solution). The angular distributions are from the
binning of Harjes~\cite{harjes} --- see Fig.~14 for the Behrend {\it et al.} binning.}}
\end{center}
\end{figure}

\begin{figure}[p]
\begin{center}
\vspace{-6cm}
~\epsfig{file=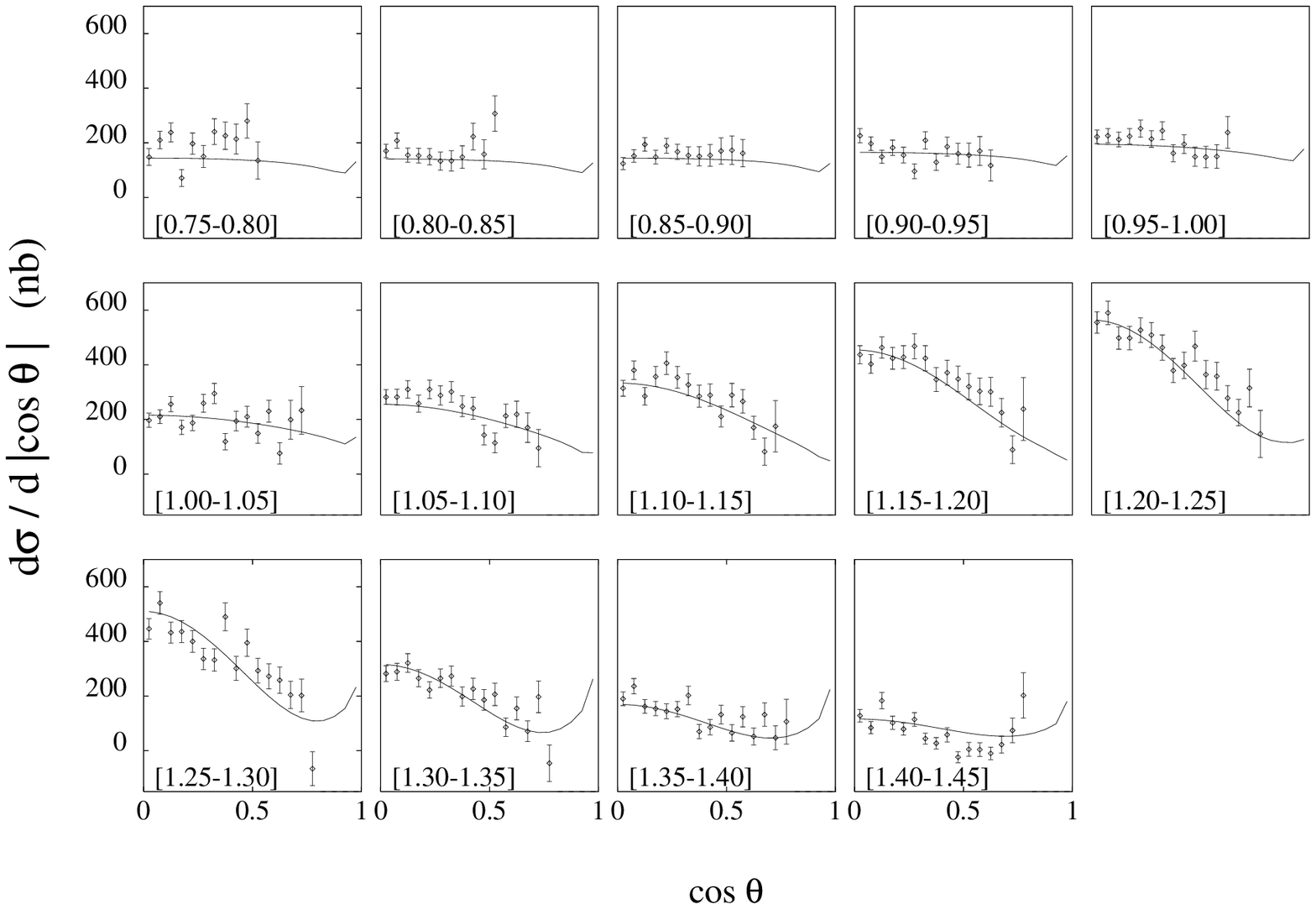,angle=0,width=13cm}
\vspace{-4.cm}
\caption{\leftskip 1cm\rightskip 1cm{The 
angular distributions as a function of $\cos\ \theta$, 
for the $\gamma \gamma \rightarrow \pi^+ \pi^-$ process,
in different $\pi\pi$ mass bins from the Behrend {\it et al.} analysis~\cite{beh} of CELLO
experiment for the {\it peak} solution, to be compared with
Figs.~9, 12, where the bins in $\cos\theta$ are larger.}}
\end{center}

\begin{center}
\vspace{-5cm}
~\epsfig{file=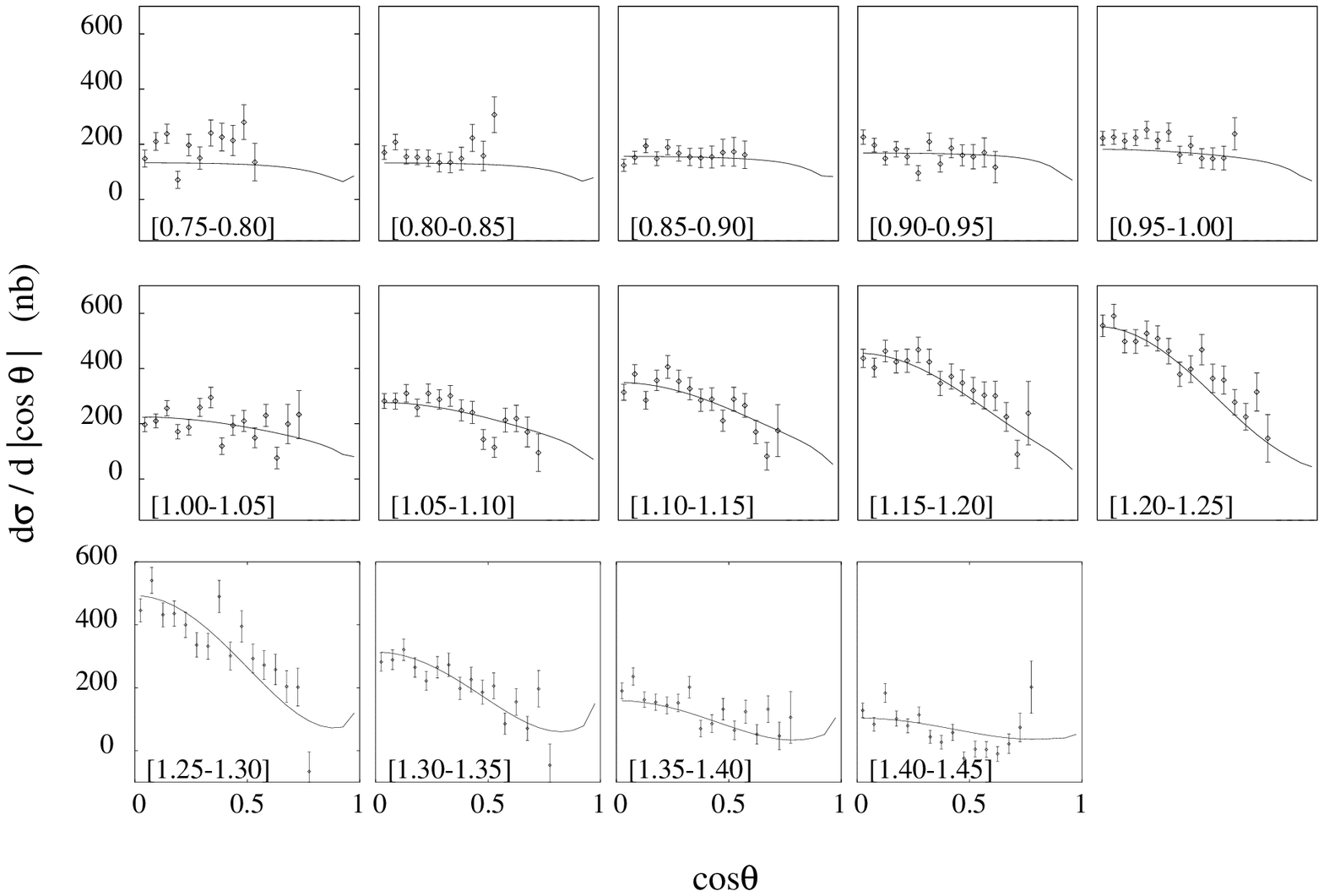,angle=0,width=13cm}
\vspace{-4.cm}
\caption{\leftskip 1cm\rightskip 1cm{ As for Fig.~13, but showing the {\it dip}
solution.}}
\end{center}
\end{figure}

\noindent To explore the neighbourhood of these best fits, it
is convenient to characterise these classes of solutions by the relative 
amount of $I=0$ $S$, $D_0$ and $D_2$ contribution to the cross-section 
$\sigma = \sigma _S + \sigma _{D_0} + \sigma _{D_2}$
at the $f_2(1270)$ peak. Each solution corresponds to a point in the 
equilateral 
triangle of height 1 with sides $\sigma _S/\sigma = 0$, 
$\sigma _{D_0}/\sigma= 0$  and $\sigma _{D_2}/\sigma = 0$ (see Figs.~15,~16). For the two 
classes of solutions, {\it peak}  and {\it dip}, we display 
the overall $\chi ^2$ per degree of freedom found for each fit in these 
equilateral triangles. 

\baselineskip=6.8mm

\noindent These diagrams, Figs.~15,~16, clearly show that the fit singles out a 
well defined region in the parameter space for each class. 
If we tried to drive them outside this area, their $\chi ^2$
values would increase very rapidly. Each rectangular flag in Figs.~15,~16 indicates one
solution is found at that particular point, and the
number which labels it is the corresponding $\chi ^2$.
The round labels indicate the position of solutions $A$ (the typical 
\lq\lq good\rq\rq ~solution) and $B$ 
(technically the \lq\lq best\rq\rq ~solution) found by Morgan and
Pennington in their previous $\gamma \gamma \to \pi \pi$ data analysis 
\cite{morpen90}. Notice how, while their favoured solution $A$, falls inside 
the
region determined by both our classes of solutions, their best solution $B$,
having very large $S$ and $D_0$ wave contributions at the
$f_2(1270)$ peak, lies considerably far away from it.

\begin{figure}[p]
\begin{center}
\vspace{-1cm}
~\epsfig{file=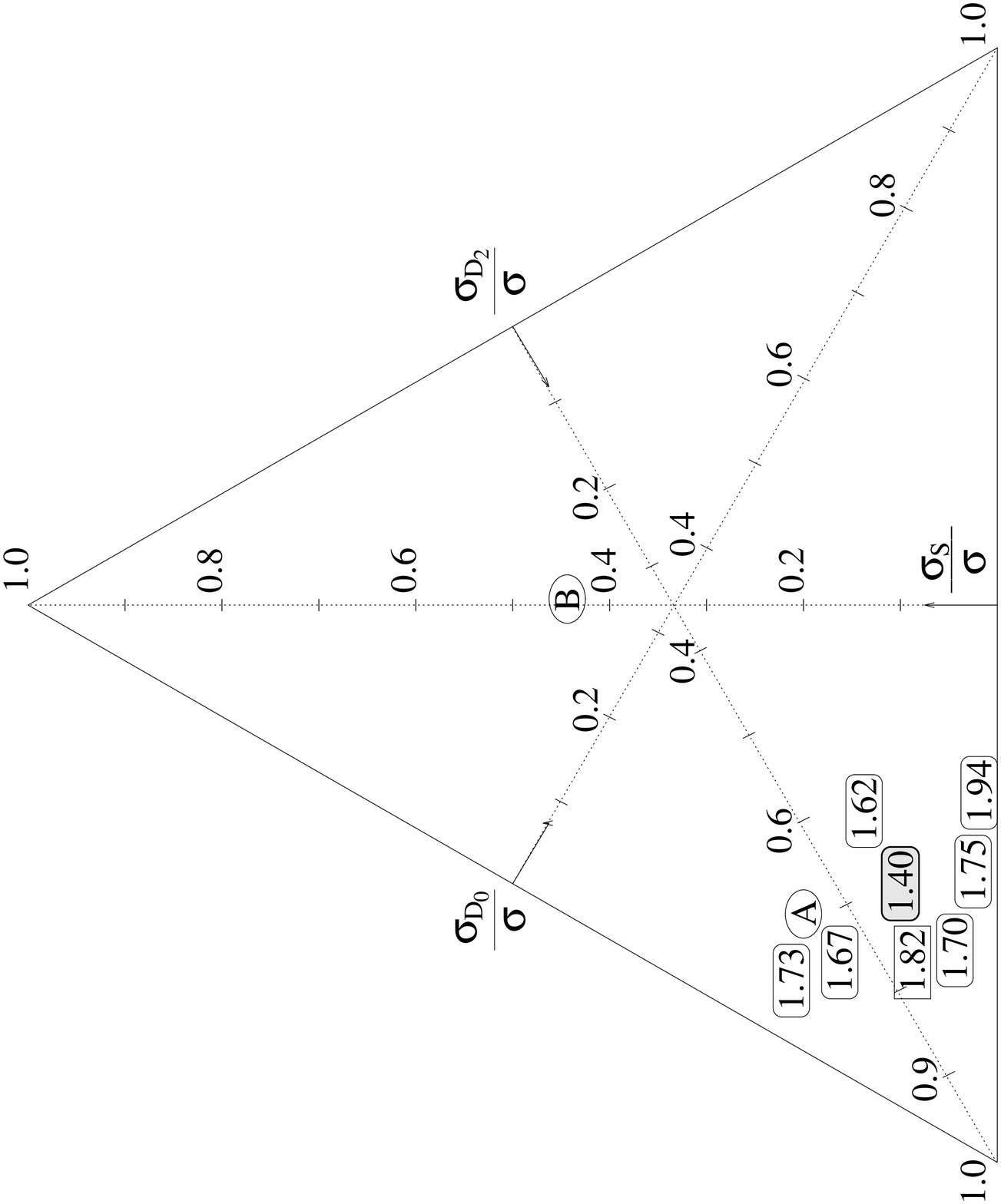,angle=-90,width=12.2cm}
\vspace{4mm}
\caption{\leftskip 1cm\rightskip 1cm{Mapping of an assortment of solutions 
with different proportions of $\sigma _S$, $\sigma _{D_0}$ and 
$\sigma _{D_2}$ at 1270 MeV labelled by the $\chi^2$ per dof
for the {\it peak} class.}}
\end{center}

\begin{center}
\vspace{-3mm}
~\epsfig{file=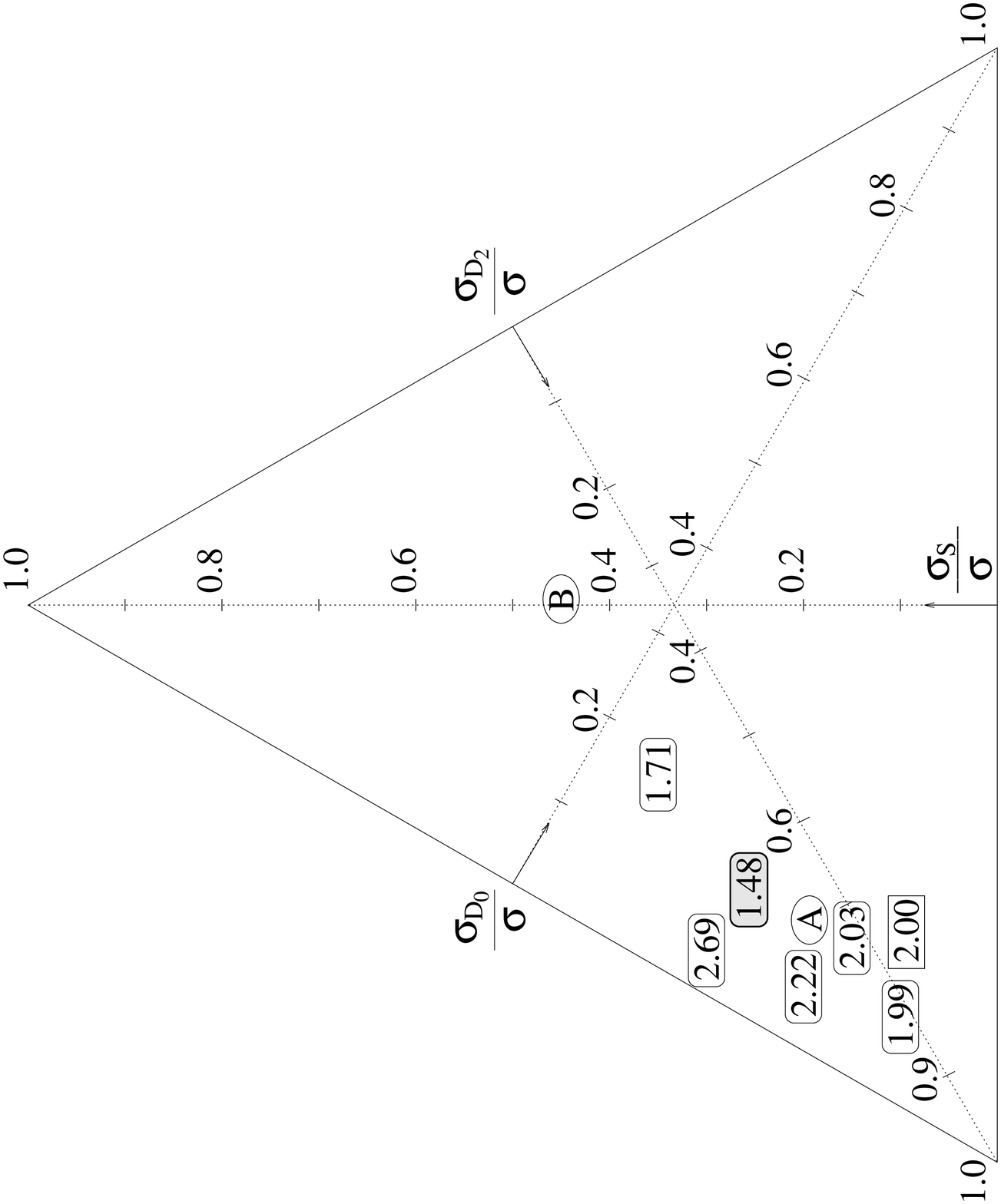,angle=-90,width=12.2cm}
\vspace{4mm}
\caption{\leftskip 1cm\rightskip 1cm{Mapping of an assortment of solutions 
with different proportions of $\sigma _S$, $\sigma _{D_0}$ and 
$\sigma _{D_2}$ at 1270 MeV labelled by the $\chi^2$ per dof
for the {\it dip} class.}}
\end{center}
\end{figure}

\noindent 
If we now compare the diagrams in Figs.~15 and~16, we see that our {\it peak}
class of solutions singles out a region in the parameter space where the
$\sigma _S/\sigma$ components are remarkably small, approximately bounded by
\ba
0.00 & < \; \sigma _S/\sigma \; < & 0.25 \;,\nn \\
0.06 & < \; \sigma _{D_0}/\sigma \; < & 0.35 \;.
\ea  
Alternatively, the {\it dip} class of solutions determine a region the boundaries
of which are given by higher values of $\sigma _S/\sigma$ and lower values
of $\sigma _{D_0}/\sigma$:
\ba
0.11 & < \; \sigma _S/\sigma \; < & 0.35 \;, \nn \\
0.03 & < \; \sigma _{D_0}/\sigma \; < & 0.16 \;.
\ea

\noindent
Furthermore, in Tables~4~and~5 we report the $\chi ^2$'s of two 
representative solutions from the {\it peak} and {\it dip} classes, to 
be compared with those corresponding to our two favoured solutions, 
respectively, shown in Tables~2~and~3. 
They have been picked out from the solutions lying in 
the regions of the parameter space shown in Figs.~15~and~16 in order to
show the difference in the $\chi ^2$'s corresponding to each data set for
an ordinary solution in either the {\it peak} or the {\it dip} class compared to our
best solution (indicated by the shaded flag).
It is immediately evident how the $\chi ^2$'s corresponding to the CELLO
experimental data, not only the overall averaged one but also the
individual $\chi ^2$ for the integrated cross section and the angular
distribution, hardly change at all. On the contrary, big variations occur
for the other data-sets, the $\chi ^2$ of which in some cases increase by a
factor of $2$. Indeed, it is all the data-sets together that determine the
features of the solutions, but we once again want to stress the point that
while a remarkably good agreement with the CELLO data is  
easily achieved  most of the time, 
the Crystal Ball and Mark II data are always hard to satisfy simultaneously.
In fact, as seen from Figs.~9~and~12, the CELLO data have finer energy
and angular bins. Consequently, it is these data that most powerfully constrain
our solutions. Thus the solutions in Tables~2--5 all have very similar
$\chi^2$ for this sector, and it is in the contributions to $\chi ^2$ from
the Mark II and Crystal Ball data-sets that they differ.
\noindent
A final comment ought to be made on the contribution to the total $\chi ^2$
from the low energy region constraints. As we have seen in Fig.~3, the
agreement of the fit solutions to the low energy amplitudes 
${\cal F}^{I=0} _{J\lambda}$, calculated by
a dispersion relation over the right--hand cut, is pretty good. The ratio
of the low energy $\chi ^2$, see Eq.~(\ref{low-en-x}), 
to the total overall $\chi ^2$ is the following
\be
\,\;{\chi ^2 _{\rm low \; energy}} / {\chi ^2 _{\rm tot}}\, =\, 0.22\qquad \;\; 
{\rm for \; the \; {\it peak} \; solution} \;, 
\ee
\be
{\chi ^2 _{\rm low \; energy}} / {\chi ^2 _{\rm tot}}\, =\,  0.17\qquad \;\; 
{\rm for \; the \; {\it dip} \; solution} \;.
\ee

\begin{table}[h]
\vspace{-0.2cm}
\begin{center}
\begin{tabular}{||c||c|c|c||}
\hline \hline
\multicolumn{4}{||c||}{\rule[-0.3cm]{0cm}{10mm} Illustrative solution no.~1
\parbox{0.8cm}{~~} $\chi ^2 _{tot} = 1.82$} \\
\hline
\rule[-0.3cm]{0cm}{10mm} Experiment &  
$\chi ^2_{{\rm average}}$ & Int. X-sect.  & Ang. distr. \\ 
\hline \hline
\rule[-0.3cm]{0cm}{10mm} 
Mark II & 2.82 & 3.21 & 2.32 \\  \hline 
\rule[-0.3cm]{0cm}{10mm}   
Cr. Ball & 2.30 & 3.09 & 2.03 \\ \hline
\rule[-0.3cm]{0cm}{10mm}   
CELLO & 1.30 & 0.64 & 1.35 \\ \hline \hline 
\end{tabular}
\vspace{0.2cm}
\caption{\leftskip 1cm\rightskip 1cm{Summary of contributions from each experiment to the total $\chi
^2$ for an illustrative solution in the {\it peak} class (see Fig.~15) to be compared with the best solution in that class. See
text for comments.}}
\end{center}

\begin{center}
\begin{tabular}{||c||c|c|c||}
\hline \hline
\multicolumn{4}{||c||}{\rule[-0.3cm]{0cm}{10mm} Illustrative solution no.~2
\parbox{0.8cm}{~~} $\chi ^2 _{tot} = 2.00$} \\
\hline
\rule[-0.3cm]{0cm}{10mm} Experiment &  
$\chi ^2_{{\rm average}}$ & Int. X-sect.  & Ang. distr. \\ 
\hline \hline
\rule[-0.3cm]{0cm}{10mm} 
Mark II & 2.87 & 2.87 & 2.87 \\  \hline 
\rule[-0.3cm]{0cm}{10mm}   
Cr. Ball & 2.90 & 3.71 & 2.64 \\ \hline
\rule[-0.3cm]{0cm}{10mm}   
CELLO & 1.42 & 0.63 & 1.49 \\ \hline \hline 
\end{tabular}
\vspace{0.2cm}
\caption{\leftskip 1cm\rightskip 1cm{Summary of contributions from each experiment to the total $\chi
^2$ for an illustrative solution in the {\it dip} class (see Fig.~16) to be compared with the best solution in that class. See
text for comments.}}
\end{center}
\end{table}

\begin{figure}[p]
\begin{center}
\vspace{-0.7cm}
~\epsfig{file=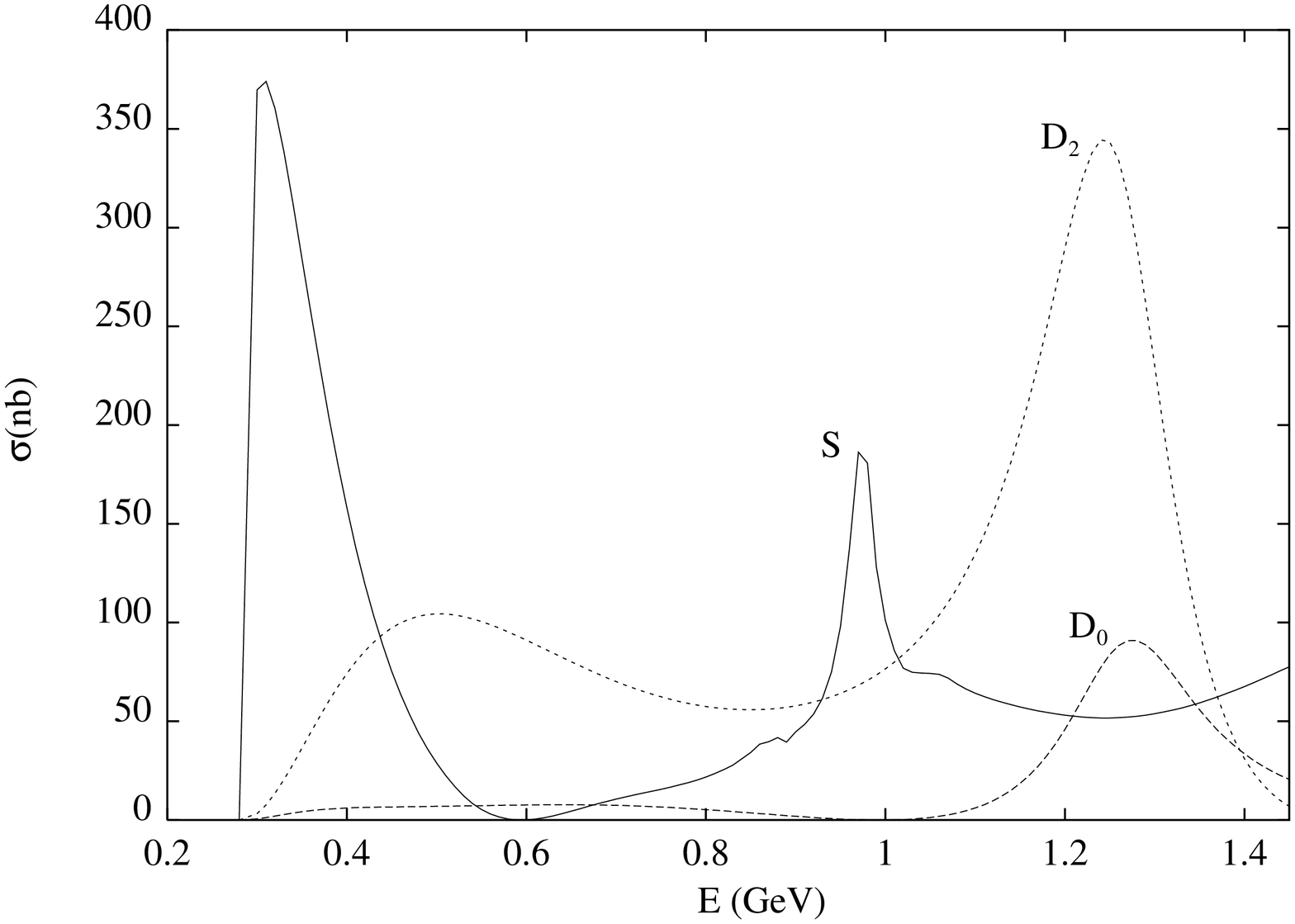,angle=-90,width=14cm}
\caption{\leftskip 0.8cm\rightskip 0.4cm{$I=0$ $\gamma \gamma \to \pi \pi$ cross-section as a function of 
energy ({\it peak} solution). Notice the two pronounced peaks in the $S$
partial wave: the first just above threshold, where it dominates the 
$\pi ^{+}\pi ^{-}$ cross--section, and the second corresponding to the
$f_0(980)$ resonance.}}
\end{center}

\begin{center}
~\epsfig{file=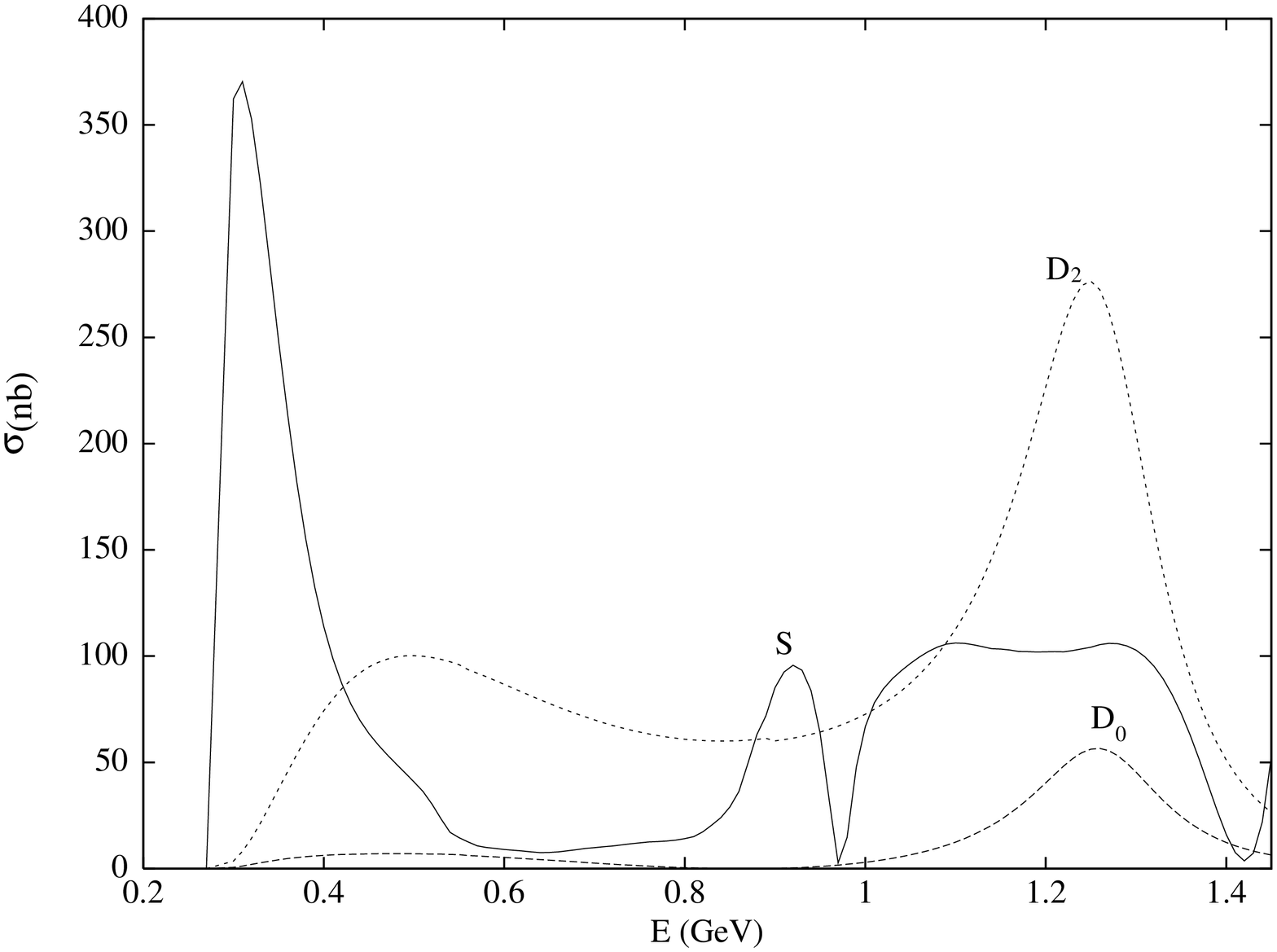,angle=-90,width=14cm}
\caption{\leftskip 0.8cm\rightskip 0.4cm{$I=0$ $\gamma \gamma \to \pi \pi$ cross-section as a function of
energy ({\it dip} solution). Notice the pronounced peak in the $S$
partial wave at threshold, where it dominates the
$\pi ^{+}\pi ^{-}$ cross--section, and the dip corresponding to the
$f_0(980)$ resonance.}} 
\end{center}
\end{figure}
\newpage
\baselineskip=7.0mm
\section{Solutions}

\noindent
Figs. 17 and 18 show the dominant partial wave components of the $I=0$ 
$\gamma \gamma \to \pi \pi$ cross-section for our two favoured solutions. 
These fall into two classes characterised by having either a peak or a dip 
in the $S$-wave cross-section in the 1 GeV region, corresponding to two 
distinct coupling patterns for the $f_0(980)$. In the {\it peak} solution 
$\alpha _K (s)$ is larger than 
$\alpha _{\pi} (s)$ in the 1 GeV region so that, in the decomposition given 
by Eq.~(7), the contribution of the hadronic reaction  
$\pi \pi \to K \overline K$ dominates over $\pi \pi \to \pi \pi$  
(see Figs.~17 and 2). As a consequence the $f_0(980)$ has a larger coupling to 
$\gamma \gamma \to K \overline K$ in the {\it peak} solution.
However, there is nevertheless a crucial contribution from $\alpha_{\pi}(s)$, 
and hence of the $\pi \pi \to \pi \pi$ amplitude, which results in the 
$f_0(980)$ peak in this $\gamma \gamma $ amplitude (Fig.~17) being sharper 
than that seen in $\pi\pi\to K \overline K$ (Fig.~2).
In the alternative class of solutions, $\alpha _{\pi}^S (s)$ dominates over 
$\alpha _K^S (s)$ in the 1 GeV region and as is clear from Fig. 18 the 
$f_0(980)$ then manifest itself as a dip as in the $\pi \pi \to \pi \pi$  
cross-section (see Fig.~2).

\noindent
In the energy region above 1 GeV, the cross section is dominated by the 
$D_2$-wave embodied in the $f_2(1270)$ resonance. An oversimplified 
description of individual channels
 is to ascribe the peak in this region wholly to $f_2(1270)$ formation in the 
 $\lambda=2$ state~\cite{hel2}. 
 Such a simplification is often used when trying to extract a $\gamma \gamma$ 
 width for the tensor mesons from a single charge  final state with limited angular 
 coverage, when a true amplitude analysis is not possible.
 Here, as in the earlier analysis by Morgan and Pennington \cite{morpen90}, 
 sizable contributions of both $S$ and $D_0$ waves in this region are strongly 
 favoured. 
 This is in good agreement with solution A of Ref. \cite{morpen90}: however 
 we do not find such large $S$ and $D_0$ contribution as their ``technically 
 best'' solution, B. We report here $\sigma_S/\sigma$ and 
$\sigma_{D_0}/\sigma$ ratios for the $I=0$ cross-sections at 1270 MeV for the
best solutions in each class:
\be
\left.
\begin{array}{r}
\rule[-0.2cm]{0cm}{9mm} 
\sigma_S/\sigma \,=\, 0.13 \\
\rule[-0.5cm]{0cm}{12mm} 
\sigma_{D_0}/\sigma \,=\, 0.22
\end{array}
\; \right\} \; {\rm for \; the \; {\it peak} \; solution\;,}
\ee  
\be
\left.
\begin{array}{r}
\rule[-0.2cm]{0cm}{9mm} 
\sigma_S/\sigma \,=\, 0.25 \\
\rule[-0.5cm]{0cm}{12mm} 
\sigma_{D_0}/\sigma \,=\, 0.13
\end{array}
\; \right\} \; {\rm for \; the \; {\it dip} \; solution\;.}
\ee  
Notice in Figs.~17,~18 how the influence of the Born term at low energies 
(modified of course by crucial final state interactions) means that even 
outside the $f_0(980)$ region the $S$-wave cross-section is not simply 
describable by a 
Breit-Wigner resonance at 600-1200 MeV, rather its contribution is spread over 
a wide region.

\baselineskip=7.2mm
\subsection{$\gamma\gamma$ couplings}

\noindent
We now calculate the $\gamma\gamma$ couplings of the resonant states in the 
threshold to $1.4$ GeV region that our amplitude analysis determines. We do 
this by two different methods~\cite{morpen90}. The first is based on the analytic continuation 
of the amplitudes we have found in our fit into the complex $s$-plane to the 
pole position. This is the only formally correct way of deducing the couplings 
of any resonance and its outcomes are free from background contamination. 
The second is a more naive approach based on the Breit-Wigner-like peak height. 
For the $f_2(1270)$, these two methods give nearly identical results, as expected 
for an uncomplicated and isolated resonance with a relatively nearby second sheet 
pole. For the $f_0(980)$ only the pole method is applicable because of the 
overlapping of this state with $K \overline K$ threshold and the broader 
$f_0(400-1200)$. For this latter broad state, only the \lq\lq peak height" provides a 
sensible measure of its $\gamma\gamma$ width, since its pole is too far from the 
real axis to be reliably located under the approximations needed to perform 
the analytic continuation, as will be clear from what follows.
 
\noindent To work out the pole residue based definition of the radiative widths, we
suppose the strong interaction amplitudes ${\cal T}_J(s)$ and the
corresponding $\gamma \gamma \to \pi \pi$ amplitudes ${\cal F}(s)$ to be
dominated by their pole contribution near the resonance pole, i.e. for
$s \sim s_R$; then we can write them in the form
\ba
\label{resT}
&{\cal T}_J(\pi \pi \to \pi \pi) (s\sim s_R)& =\;\; \frac{g ^2 _{\pi}}{s_R-s} \;,
\\
\nonumber \\
\label{resF}
&{\cal F}_{J\lambda}(\gamma \gamma \to \pi \pi) (s\sim s_R)& =\;\;
\frac{g_{\gamma}\, g _{\pi}}{s_R-s}\;.
\ea
It is easy to see that the couplings $g _{\pi}$ and $g_{\gamma}$ can be
extracted as the residues of these amplitudes at the resonance pole $s_R$.
Thanks to the parametrization of the amplitudes ${\cal T}$ in terms of the
$K$--matrix elements, as given in Eq.~(10), 
we know their numerator, $N_{\pi}(s)$ and $N_{K}(s)$ respectively, and 
denominator, $D(s)$, which is the same for the two of them. 
So we can use the expressions
\be
{\cal T}_J(\pi \pi \to \pi \pi)\; =\; \frac{N_{\pi}(s)}{D(s)}\;,
\label{ND-Tp}
\ee
\be
{\cal T}_J(\pi \pi \to K \overline K)\; =\; \frac{N_{K}(s)}{D(s)}\;,
\label{ND-TK}
\ee
and Eq.~(7), to write ${\cal F}_{J\lambda}(\gamma \gamma \to \pi \pi)$ as:
\be
{\cal F}_{J\lambda}(\gamma \gamma \to \pi \pi)\, =\, 
\frac{\overline \alpha ^{J\lambda}_{\pi} (s) \, N_{\pi}(s) + 
\overline \alpha ^{J\lambda}_{K} (s) \, N_{K}(s)}{D(s)}\;.
\ee
Now, in the region nearby $s_R$ we can make a Taylor expansion of the
function $D(s)$, truncated at the first order
\be
D(s\sim s_R)\simeq D(s_R)+D'(s_R)(s-s_R)\,=\, D'(s_R)(s-s_R)\;,
\ee
where by definition $D(s_R) = 0$ at the resonance pole. 
Finally, by substituting this into Eqs.~(\ref{ND-Tp},\ref{ND-TK})
and comparing with Eqs.~(\ref{resT},\ref{resF}),
we find
\be
g_{\pi}^2\,=\,\frac{N_{\pi}(s_R)}{D'(s_R)}\;,
\ee  
and
\be
g_{\gamma}\,g_{\pi}\, =\, \frac{\overline \alpha ^{J\lambda}_{\pi} (s_R) \, 
N_{\pi}(s_R) + 
\overline \alpha ^{J\lambda}_{K} (s_R) \, N_{K}(s_R)}{D'(s_R)}\;,
\ee
from which the coupling $g_{\gamma}$ can readily be calculated.
The corresponding width is then evaluated using the formula
\be
\Gamma ^R _{\gamma \gamma} ({\rm pole})\, =\, \frac{\alpha ^2\, \beta_R\; |g_{\gamma}|^2}
{ 4(2J+1) m_R} \; ,
\label{Gamma-pole}
\ee
where $\alpha=1/137$ is the fine structure constant, and $\beta_R =
(1-4m_{\pi} ^2 /m_R^2)^{1/2} \simeq 1$.

\noindent
Because of its very large width, the $f_0(400-1200)$ coupling to two photons
cannot be calculated with this technique, since one cannot reliably
continue so far into the complex plane.
As an alternative, we give a rough estimate of its 
width by using an expression based on the standard resonance peak formula
\be
\Gamma ^R _{\gamma \gamma}({\rm peak})\,=\,\frac{\sigma _{\gamma \gamma}
({\rm res. \; peak})\,m^2_R\, \Gamma _{tot}}{8\pi (\hbar c)^2 (2J+1)BR}\;,
\label{Gamma-phys}
\ee
where $BR$ is the hadronic branching ratio for the final state considered.  

\begin{table}[t]
\begin{center}
\begin{tabular}{|l||c|c|c|}
\cline{2-4}
\multicolumn{1}{c|}{~~}
&\rule[-0.6cm]{0cm}{8mm} $\Gamma (f_2(1270))$ 
& $\Gamma (f_0(980))$ & $\Gamma (f_0(400-1200))$
\rule[0.4cm]{0cm}{8mm}  \\  
\hline \hline 
\rule[-0.4cm]{0cm}{8mm}   
Peak solution  & 3.04 & ~~0.13-0.36~~ & 3.0 
\rule[0.4cm]{0cm}{8mm}   \\ \hline
\rule[-0.4cm]{0cm}{8mm} 
Dip solution  & 2.64 & 0.32     & 4.7 
\rule[0.4cm]{0cm}{8mm}   \\
\hline 
\end{tabular}
\vspace{0.4cm}
\caption{\leftskip 2.cm\rightskip 2.cm{Two photon partial widths in $keV$
for the meson states in our two 
classes of~solutions.}}
\end{center}
\vspace{-3mm}
\end{table}

\noindent
Table~6 shows the results we obtain for the $\gamma \gamma$ 
widths of the $f_0(980)$,
$f_0(400-1200)$, $f_2(1270)$ from either the pole or the peak--height definition, as
appropriate, when choosing solution~1 and solution~2, respectively. 

\noindent
The continuation to the second sheet pole for the $f_0(980)$ is rather 
sensitive to the parametrization of the $K$--matrix (and hence of the $T$--matrix, 
Eq.~10) used.
In the ReVAMP analysis~\cite{revamp} described in Sect.~2,
the $K$--matrix elements are  given by a single  pole plus different 
order polynomials. These each fit the hadronic data equally well. For the 
{\it dip} solution, the $\gamma\gamma$ width for the $f_0(980)$ is $0.31$ keV 
for  ReVAMP1, and 
$0.34$ keV for ReVAMP2, so rather little change. However, 
for the peak solution for which there is a much stronger interplay between 
the ${\cal T}_S(\pi \pi \to \pi \pi)$ and the 
${\cal T}_S(\pi \pi \to K \overline K)$ contributions to Eq.~(7), 
we find a $\gamma\gamma$ width of $0.13$ keV for ReVAMP1 and $0.36$ keV for 
ReVAMP2. Hence the values in 
Table~6.

\noindent 
From the variation between different solutions within the classes 
indicated in Figs.~15 and 16, we estimate that the uncertainty on the 
$\gamma\gamma$ width of the $f_2(1270)$ is $\pm 0.08$ keV within each class of 
solutions. To this must be added 
a $5\%$ uncertainty in the absolute normalization. Consequently, it is the 
difference between solutions (rather than within a given class) that 
constitutes the major uncertainty, and so we quote 
\be
\Gamma(f_2(1270) \to \gamma\gamma)\, =\, (2.84\,\pm\, 0.35) \; {\rm keV}\; .
\ee
Whilst this is in good agreement with the PDG'98 {\it estimated} value of $(2.8\pm 0.4)$ keV, it  is somewhat at variance with the PDG'98 {\it fitted}
value of $\left(2.44^{+0.32}_{-0.29}\right)$ keV,
 based on 
several different analyses~\cite{PDG} of either $\pi ^+ \pi ^-$ or $\pi ^0 \pi ^0$ 
data separately using quite different assumptions. 
It is important to stress that our value is nearest 
to a model independent amplitude analysis result one can presently achieve.

\noindent
The uncertanties on the $\gamma\gamma$ widths of the $f_0(980)$ and 
$f_0(400-1200)$ are more problematic. For the $f_0(980)$ we quote
\be
\Gamma (f_0(980)\to \gamma\gamma)\, =\, (0.28\, ^{+0.09}_{-0.13}) \; {\rm keV}.
\ee
However, a decision on whether the {\it dip} or {\it peak} solution was correct 
would reduce the uncertainty dramatically. 

\noindent
For the $f_0(400-1200)$ the estimate is much cruder and a $50\%$ uncertainty 
is likely, giving
\be
\Gamma (f_0(400-1200)\to \gamma\gamma)\, =\, (3.8\, \pm\, 1.5) \; {\rm keV}.
\ee
Once again discriminating between the classes of solutions would change the central 
value and reduce the error within these ranges.

\noindent
What about the composition of these states ?
Quite independently, we, of course, know the $f_2(1270)$ is the 
$n\overline n$ member of an ideally mixed multiplet. The $\gamma\gamma$ widths 
of the neutral members are then expected to be in the ratio of the squares of 
the average squared charges of their constituents, so
\be
\Gamma (f_2\to \gamma\gamma) : \Gamma (a_2\to \gamma\gamma) :
\Gamma (f_2'\to \gamma\gamma)\, =\, 25:9:2 \; .
\ee

\noindent Experiments give ratios very close to this.
The prediction for the $\gamma\gamma$ width not only depends on the charges of 
the constituents to which the photons couple, but to the probability that 
these constituents annihilate --- their overlap. For members of a $q\overline q$ 
multiplet we expect these probabilities should be roughly equal. Experiment for 
the lowest tensor nonet confirms this. (Any differences can readily be explained by a small departure from ideal mixing --- see the Appendix of
Ref.~\cite{morpen90}.) 

\noindent For the scalars, we need a certain modelling.  The simplest would
be to assume the lightest scalars are the shadow of the tensor nonet. Then non-relativistically,
Chanowitz~\cite{chano} deduced that the corresponding tensor and scalar two
photon widths are related by
\be
\Gamma (0^{++} \to \gamma \gamma)\, =\, \left( \frac{15}{4} \right) \times 
\left( \frac{m_0}{m_2} \right)^3 \times \Gamma (2^{++} \to \gamma \gamma)\;,
\ee
\noindent where the factor of $(m_0/m_2)^3$ takes account of the mass splitting. This gives
the predictions in Table~7, under the assumption that the $f_0$ states are either purely
$s{\overline s}$ or non-strange $n{\overline n}$. Relativistic  corrections
to Eq.~(56) have been computed~\cite{lirel} and found to reduce
the ratio $\Gamma(0^{++}\to\gamma\gamma)/\Gamma(2^{++}\to\gamma\gamma)$ by 
as much as a factor of 2 for light quark systems.  An oft-cited alternative
structure for the $f_0(980)$ is as a $K{\overline K}$ molecule~\cite{wisgur}.
Then though the
fourth power of the charge of the constituents is far greater for
a $K{\overline K}$ molecule than for a simple $s{\overline s}$ bound state, the molecule is a much more diffuse
system, so that the probability of the kaons annihilating to form photons is 
strongly suppressed.  Thus in the non-relativistic potential model
of Weinstein and Isgur~\cite{wisgur}, computation of the molecular radiative width gives 
 $\Gamma(f_0(K{\overline K}\to\gamma\gamma)) = 0.6$ keV~\cite{barnesgg}.
\begin{table}[h]
\begin{center}
\begin{tabular}{|l||c|c|c|}
\cline{2-4}
\multicolumn{1}{c|}{~~}
&\rule[-0.6cm]{0cm}{8mm}  $n \overline n$
& $s \overline s$ & 
$K \overline K$ 
\rule[0.4cm]{0cm}{8mm}  \\  
\hline \hline 
\rule[-0.4cm]{0cm}{8mm}   
$\Gamma _{0^{++}}$ & ~~~4.5~~~ & ~~~0.4~~~ & ~~~0.6~~~  
\rule[0.4cm]{0cm}{8mm}   \\
\hline 
\end{tabular}
\vspace{0.4cm}
\caption{\leftskip 2.5cm\rightskip 2.5cm{Two photon partial widths in $keV$ predicted for a
conventional $q \overline q$ nonet of 
scalar states and for a $K \overline K$ molecule.}}
\end{center}
\vspace{-4mm}
\end{table}

\noindent Even this may be far too simplistic for these $0^{++}$ states.
From the work of Tornqvist~\cite{nils} (replicated by one of us~\cite{elena}), we know the dressing of bare $q{\overline q}$ states, by the hadrons into which the physically observed mesons decay, is particularly large for scalars. In lattice-speak, their {\it unquenching} is a big effect. In a scheme where it's assumed
 pseudoscalar meson pairs provide the dominant dressing,
 the $f_0(980)$ has not only an
$s{\overline s}$ (and smaller $n{\overline n}$) component, but a large
$K{\overline K}$ admixture, too. It is not a case of the physical hadron being either a
$q{\overline q}$ bound state  or a $K{\overline K}$ molecule, but it is in fact both!
What this non-perturbative treatment predicts for the radiative width of the $f_0(980)$ is a calculation under way.

\noindent We know there are other scalars beyond the region of our analysis:
the $f_0(1500)$ and $f_J(1710)$ in particular~\cite{PDG}. Some claim there is also an
$f_0(1370)$. In the past, we have argued that this may just be the higher end of the broad $f_0(400-1200)$, cut-off below by $K{\overline K}$ and $4\pi$ phase-space. Interestingly, the very recent potential model analysis of Kaminski {\it et al.}~\cite{kam}
suggests the $f_0(1370)$ may actually be the same as the $f_0(1500)$.
In any event, the radiative width of all these states is a key pointer to their composition, gluonic or otherwise~\cite{glue,lund}. 
At present, only crude estimates are possible, for instance~\cite{aleph} that
$\Gamma(f_0(1500)\to\gamma\gamma) < 0.17$ keV. To achieve something more
is the challenge for the future. A major task is to extend the present Amplitude Analysis beyond 1.4 GeV. This requires
a study of two photon production of not just two pion final states, but
$4\pi$ and $K\overline K$ too. Only by a  detailed analysis
of these final states simultaneously, can  we hope to extract a true scalar signal from under the
dominant spin 2 effects in the region from 1.3--1.8 GeV and so deduce the
radiative widths of the $f_0(1500)$ and $f_J(1710)$.

\noindent
We have seen that presently published data allow  two classes of solutions,
distinguished by the way the $f_0(980)$ couples to $\gamma\gamma$. A primary 
aim must be to distinguish between these. We believe that data with sufficient precision may already
have been taken
at CLEO that could do this~\cite{paar}.  However, these results have not yet been corrected for acceptance and efficiency, and sadly may not be. To go to higher masses
within the resonance region may well be possible when corrected results from LEP2 and
future experiments at $B$--factories become available.  The challenge for theory is equally demanding: it
is to deduce what the radiative widths for the $f_0(980)$ and for the $f_0(400-1200)$  we have determined here
from experiment, and summarised
in Table~6,
tell us about the underlying nature of these dressed hadrons.
Only then may we hope to solve the enigma of the scalars: states that are
intimately related to the breakdown of chiral symmetry and hence reflect the very nature of the QCD vacuum.

\vspace{5mm} 

\section*{Acknowledgements}
Preliminary work on this analysis began when one of us (MRP) visited
DESY to have discussions with Dr K. Karch and Prof. J.H. Bienlein.
Without their initial input and interest, this Amplitude Analysis would not have come to fruition: for this we are most grateful.
We  acknowledge partial  support from the EU-TMR Programme, Contract No. FRMX-T96-0008.

\end{document}